\documentclass[a4paper,12pt]{article}

\usepackage{amsmath}
\usepackage{amssymb}
\usepackage{slashed}
\usepackage{amsthm}
\usepackage{xcolor}
\usepackage{calc}

\numberwithin{equation}{section}
\def\appendix#1{\addtocounter{section}{1}\setcounter{equation}{0}
\renewcommand{\thesection}{\Alph{section}}
\section*{Appendix \thesection\protect\indent \parbox[t]{11.15cm}{#1}}
\addcontentsline{toc}{section}{Appendix \thesection\ \ \ #1}}

\usepackage{accents}

\newcommand{\bea}{\begin{eqnarray}}
\newcommand{\eea}{\end{eqnarray}}

\begin{document}

\begin{titlepage}
\begin{center}

\vspace*{-1.0cm}

\hfill  DMUS-MP-22-04
\\
\vspace{2.0cm}

\renewcommand{\thefootnote}{\fnsymbol{footnote}}
{\Large{\bf Heterotic de-Sitter Solutions}}
\vskip1cm
\vskip 1.3cm
D. Farotti
\vskip 1cm
{\small{\it
Department of Mathematics,
University of Surrey \\
Guildford, GU2 7XH, UK.}\\
\texttt{d.farotti@surrey.ac.uk}}

\end{center}
\bigskip
\begin{center}
{\bf Abstract}
\end{center}
We classify all warped product de-Sitter backgrounds in heterotic supergravity, up to two loops. We find that warped $dS_n$ backgrounds, with $n\ge 3$, are $\mathbb{R}^{1,n}\times M_{9-n}$, where $M_{9-n}$ is a $(9-n)$-dimensional Riemannian manifold. Moreover, we establish that warped $dS_2$ backgrounds are $AdS_3\times M_7$, where $M_7$ is a 7-dimensional Riemannian manifold.

\end{titlepage}

\section{Introduction}

Cosmological observations in the nineties revealed that at the present time the universe possesses a small positive cosmological constant \cite{SupernovaCosmologyProject:1998vns}. Moreover, shortly after the big bang, there is good reason to think that the universe underwent a period of exponential expansion, called inflation \cite{Guth:1980zm}. Hence de-Sitter space seems to play an important role in the understanding of our past and present universe. Within the framework of string theory, a quantum gravity description of de-Sitter space presents various difficulties \cite{Witten:2001kn}. Moreover, in the classical limit, there are no go-theorems stating that there are no smooth compactifications of ten or eleven-dimensional supergravity to warped de-Sitter space of any dimension \cite{gbds, deWit:1986mwo, Maldacena:2000mw}. Taking these issues into account, it has been shown that there exist vacua with positive cosmological constant within the string landscape \cite{Kachru:2003aw}. However, these vacua are meta-stable and decay to the true vacuum \cite{Goheer:2002vf}. In the context of holography, one of the most notable proposed dual descriptions of de-Sitter space is the dS-CFT correspondence, which directly tries to generalize the original AdS-CFT correspondence \cite{Maldacena:1997re} to de-Sitter space \cite{Strominger:2001pn, Strominger:2001gp, Malda2003}.  

To make progress towards the classification of de-Sitter backgrounds in string theory, heterotic supergravity is a natural place to start. Indeed, supersymmetric heterotic supergravity backgrounds have been thoroughly investigated using spinorial geometry techniques: the Killing spinor equations (KSE) have been solved in all cases and the geometry of these backgrounds has been determined \cite{spinorial1, spinorial2,spinorial3}. Moreover, the $\alpha^{'}$ corrections to heterotic backgrounds can be studied in the sigma model approach \cite{Hull}. According to this scheme, the field equations correspond to the vanishing of the sigma model beta functions, which are an infinite perturbation series in the string tension $\alpha^{'}$. In particular, the Bianchi identities are modified: the Green-Schwarz anomaly-cancellation mechanism requires that at one-loop $dH$ is equal to the difference of the Pontrjagin forms of spacetime and the vector bundle of the gauge sector \cite{Green:1984sg}. Consistency of the theory then requires to take into account corrections up to two loops in the bosonic field equations and the KSE; such corrections have a simpler form compared to type II supergravity theories \cite{intro1,intro2,intro3}, hence anomaly corrected heterotic backgrounds are more tractable. Anomaly corrected backgrounds in heterotic theory have been studied in \cite{Fontanella:2016aok}, in the context of supersymmetric near-horizon geometries and in \cite{Beck:2015gqa}, in the context of the classification of supersymmetric warped $AdS$ backgrounds.

In this paper we analyze warped de-Sitter backgrounds $dS_n\times_w M_{10-n}$ in heterotic supergravity, including $\alpha^{'}$ corrections up to two loops in sigma model perturbation theory.
In particular, we require that when $\alpha^{'}\to 0$, the geometry of the spacetime is exactly $dS_n\times_w M_{10-n}$. This corresponds to take the zero-th order term $k^0$ in the $\alpha^{'}$ expansion of the cosmological constant $k$ to be non-zero, that is $k^0=\ell ^{-2}$. As stated before, there are many no-go theorems which rule out de-Sitter solutions in supergravity; these no-go theorems assume that the warp factor and fluxes are smooth, and that the internal manifold is smooth and compact without boundary. Thus, our motivation is to construct a systematic classification of supersymmetric de Sitter solutions in supergravity theories from a purely \textit{local} perspective. This evades the no-go theorems mentioned before. In particular, in this work we do not assume the warp factor to be smooth, nor the internal manifold $M_{10-n}$ to be either smooth or compact. Our local analysis makes use only of the bosonic field equations, the Bianchi identities and the KSE of the theory. Furthermore, we do not assume the Killing spinors to factorize, as it is known that such factorizations can produce a miscounting of supersymmetries \cite{Gran:2016zxk}. We find that warped product de Sitter solutions in heterotic supergravity are very restricted. Indeed, we show that warped product $dS_2$ solutions are direct product $AdS_3$ solutions and all warped product $dS_n$ solutions for $n \ge 3$ are $\mathbb{R}^{1,n}\times M_{9-n}$. These results complement \cite{Kutasov:2015eba}, where, utilizing world-sheet CFT arguments, it is shown that in heterotic string theory $dS_n$ vacua with $n\ge 4$ are forbidden to all orders in $\alpha^{'}$. 

It is useful to compare and contrast the heterotic results to the case of $D=11$ supergravity. The warped product $dS_n$ solutions of $D = 11$ supergravity have similar foliation properties for $n\ge 5$. In particular, for $n=5$, $dS_5$ arises as a conformal foliation of $\mathbb{R}^{1,5}$, corresponding to the directions along the M5-brane worldvolume \cite{Farotti:2022xsd}. In contrast, the warped product $dS_4$ solutions of $D=11$ supergravity preserving $N = 8$ supersymmetry, which have been classified in \cite{DiGioia:2022bqg} using spinorial geometry techniques, do not have an analogous foliation into $AdS_5$ or $\mathbb{R}^{1,4}$. Moreover, for warped product $dS_3$ backgrounds in $D=11$ supergravity, we exhibit a rich algebraic structure,  which also suggests that such solutions are not foliations of $dS_3$ into $AdS_4$ or $\mathbb{R}^{1,3}$. 

The paper is organized as follows. In section 2 we present the bosonic field equations, the Bianchi identities and the KSE of heterotic perturbation theory, up to two loops. Moreover, we formulate the ansatz for the bosonic fields in the case of warped de-Sitter backgrounds $dS_n\times_w M_{10-n}$ and we reduce the 10-dimensional gravitino KSE along the internal manifold $M_{10-n}$, for generic $n$. In section 3 we analyse warped product $dS_n$ backgrounds with $n\ge 4$, up to two loops. In particular, we decompose the field equations, the Bianchi identities, the dilatino and the gaugino KSE along $M_{10-n}$ and we utilize this to show that the 10-dimensional spacetime is $\mathbb{R}^{1,n}\times M_{9-n}$, where $M_{9-n}$ is a $(9-n)$-dimensional Riemannian manifold. Moreover, we show that the 1-form $d\Phi$, the 3-form $H$ and the non abelian field strength $F$ have non-vanishing components only along $M_{9-n}$. In section 4 we study warped product $dS_3$ backgrounds, up to two loops. In this case, we show that the 10-dimensional spacetime is $\mathbb{R}^{1,3}\times M_{6}$, where $M_{6}$ is a $6$-dimensional Riemannian manifold. Moreover, the 1-form $d\Phi$, the 3-form $H$ and the non abelian field strength $F$ have non-vanishing components only along $M_6$. In section 5 we investigate warped $dS_2$ backgrounds, up to two loops. After decomposing the bosonic field equations, Bianchi identities and algebraic KSE on $M_8$, we show that all these backgrounds are $AdS_3\times M_7$, where $M_7$ is a 7-dimensional Riemannian manifold. In section 6 we study the Killing superalgebra $\mathfrak{g}$ of warped $dS_3$ backgrounds in $D=11$ supergravity and we show that there is a rich algebraic structure underlying $\mathfrak{g}$. Some brief conclusions are presented in Section 7. The paper ends with two appendices. Appendix A contains our conventions for the curvature tensor of a connection and some useful formulae which are used in the computations of the paper. Appendix B contains some details about the Killing superalgebra of $D=11$ warped $dS_3$ backgrounds. In particular, we list the spinorial Lie derivatives and the conditions arising from the super-Jacobi identities.

\section{Bosonic field equations and KSE}

In this section we present the bosonic field equations and the Killing spinor equations (KSE) of heterotic perturbation theory, up to two loops. The bosonic field equations of heterotic supergravity up to two loops in sigma model perturbation theory are given by \cite{Hull}
\begin{eqnarray}
&&R_{MN}-\frac{1}{4}H_{ML_1L_2}H_N^{~L_1L_2}+2\nabla_M\nabla_N\Phi
\nonumber \\
&&+\frac{\alpha^{'}}{4}\bigg(\check{R}_{ML_1,L_2L_3}\check{R}_N^{~~L_1,L_2L_3}-F_{MLij}F_N^{~Lij}\bigg)=\mathcal{O}(\alpha^{'2})
\label{Einsteincorr}
\end{eqnarray}
\begin{eqnarray}
\nabla^M(e^{-2\Phi}H_{MN_1N_2})=\mathcal{O}(\alpha^{'2})
\label{gaugeHcorr}
\end{eqnarray}
\begin{eqnarray}
\nabla^M(e^{-2\Phi}F_{MN})+\frac{1}{2}e^{-2\Phi}H_{NL_1L_2}F^{L_1L_2}=\mathcal{O}(\alpha^{'})
\label{gaugeFcorr}
\end{eqnarray}
\begin{eqnarray}
&&\nabla^M\nabla_M\Phi-2(\nabla\Phi)^2+\frac{1}{12}H_{MNR}H^{MNR}
\nonumber \\
&&-\frac{\alpha^{'}}{16}\bigg(\check{R}_{N_1N_2,N_3N_4}\check{R}^{N_1N_2,N_3N_4}-F_{MNij}F^{MNij}\bigg)=\mathcal{O}(\alpha^{'2})
\label{dilatoncorr}
\end{eqnarray}
where $M,N\dots$ are 10-dimensional frame indices, $i,j$ are gauge indices and $\check{R}$ is the curvature tensor with respect to the connection $\check{\nabla}:=\nabla-\frac{1}{2}H$. Our conventions for the curvature are presented in Appendix A. In the following, gauge indices are understood. Moreover, the Bianchi identities read
\begin{eqnarray}
dH=-\frac{\alpha^{'}}{4}\bigg(\textrm{tr} (\check{R}\wedge \check{R})-\textrm{tr} (F\wedge F)\bigg)+\mathcal{O}(\alpha^{'2})~.
\label{BianchiDH}
\end{eqnarray}
We remark that the right hand side of \eqref{BianchiDH} is required for anomaly cancellations and is expressed as the difference of two Pontryagin forms, one of the tangent space of spacetime and the other of the gauge sector bundle. Furthermore, the KSEs are given by
\begin{eqnarray}
\hat{\nabla}_M\epsilon:=\nabla_M\epsilon-\frac{1}{8}H_{MN_1N_2}\Gamma^{N_1N_2}\epsilon=\mathcal{O}(\alpha^{'2})
\label{KSEcorr1}
\end{eqnarray}
\begin{eqnarray}
\bigg(\Gamma^M\nabla_M\Phi-\frac{1}{12}H_{N_1N_2N_3}\Gamma^{N_1N_2N_3}\bigg)\epsilon=\mathcal{O}(\alpha^{'2})
\label{KSEcorr2}
\end{eqnarray}
\begin{eqnarray}
F_{MN}\Gamma^{MN}\epsilon=\mathcal{O}(\alpha^{'})
\label{KSEcorr3}
\end{eqnarray}
where $\hat{\nabla}:=\nabla+\frac{1}{2}H$. Notice that the KSEs have the same form up to two loops in sigma model perturbation theory as that at the zero-th order. \\
\indent
In our work, we analyze warped product $dS_n$ backgrounds in heterotic supergravity, up to two loops.  In the perturbation theory scheme, the bosonic fields and the Killing spinors are expanded in $\alpha^{'}$. The metric on the $D=10$ spacetime $M_{10}$ is given by
\begin{eqnarray}
ds^2(M_{10})=A^2ds^2(dS_n)+ds^2(M_{10-n})+\mathcal{O}(\alpha^{'2})
\label{EQ1}
\end{eqnarray}
where $A$ is a function of the co-ordinates of the Riemannian manifold $M_{10-n}$ and
\begin{eqnarray}
ds^2(dS_n)=\frac{1}{(1+\frac{k}{4}|x|^2)^2}\eta_{\mu\nu}dx^{\mu} dx^{\nu}~~~~~\mu,\nu=0,1,\dots n-1
\label{dSD}
\end{eqnarray}
is the metric of $n$-dimensional de-Sitter spacetime, where $|x|^2=\eta_{\mu\nu}x^{\mu}x^{\nu}$ and $k=\frac{1}{\ell^2}$. Notice that $k$ itself admits a perturbation series in $\alpha^{'}$, that is
\begin{eqnarray}
k=k^0+k^1\alpha^{'}+\mathcal{O}(\alpha^{'2})~.
\end{eqnarray}
In our analysis, we consider $k^0=\frac{1}{\ell^2}$, that is, in the limit $\alpha^{'}\to 0$ \eqref{EQ1} reduces to $dS_n\times_w M^{10-n}$. Let us introduce on $M_{10}$ the co-frame
\begin{eqnarray}
\textbf{e}^{\mu}=\frac{A}{\mathcal{U}}dx^{\mu}~,~~~~~\textbf{e}^a=e^a_{~\alpha}(y)dy^{\alpha}
\label{viel1}
\end{eqnarray}
where $a=n,n+1,\dots , 9$, $y^{\alpha}$ denote the co-ordinates on $M_{10-n}$ and
\begin{eqnarray}
\mathcal{U}=1+\frac{k}{4}|x|^2~.
\label{R}
\end{eqnarray}
In terms of the co-frame \eqref{viel1}, the metric tensor \eqref{EQ1} reads
\begin{eqnarray}
ds^2(M_{10})=\eta_{\mu\nu}\textbf{e}^{\mu} \textbf{e}^{\nu}+ds^2(M_{10-n})+\mathcal{O}(\alpha^{'2})
\label{11DM}
\end{eqnarray}
where
\begin{eqnarray}
ds^2(M_{10-n})=\delta_{ab}\textbf{e}^a \textbf{e}^b~.
\label{m11-n}
\end{eqnarray}
Moreover, we assume that the dilaton $\Phi$ is a function of the co-ordinates of $M_{10-n}$. The ansatz for the 3-form $H$ and the non-abelian field strength $F$ depends on $n$. Assuming $H$ and $F$ to be invariant under the isometry group of $dS_n$ up to two loops, it follows that
\begin{itemize}
\item{$n \ge 4$}
\begin{eqnarray}
H=X+\mathcal{O}(\alpha^{'2})
\label{Hdsn}
\end{eqnarray}
\begin{eqnarray}
F=\mathcal{F}+\mathcal{O}(\alpha^{'})
\label{Fdsn}
\end{eqnarray}
where $X\in\Omega^3(M_{10-n})$ and $\mathcal{F}$ is a 2-form on $M_{10-n}$ taking values in $\mathfrak{g}$, where we denote by $\mathfrak{g}$ the Lie algebra of the gauge group.

\item{$n=3$}

\begin{eqnarray}
H=f~\textrm{dvol}(dS_3)+X+\mathcal{O}(\alpha^{'2})
\label{Hds3}
\end{eqnarray}
\begin{eqnarray}
F=\mathcal{F}+\mathcal{O}(\alpha^{'})
\end{eqnarray}
where $f\in\Omega^0(M_7)$, $X\in\Omega^3(M_7)$ and $\mathcal{F}$ is a 2-form on $M_{7}$ taking values in $\mathfrak{g}$.

\item{$n=2$}

\begin{eqnarray}
H=B\wedge \textrm{dvol}(dS_2)+X+\mathcal{O}(\alpha^{'2})
\label{Hds2}
\end{eqnarray}
\begin{eqnarray}
F=f\textrm{dvol}(dS_2)+\mathcal{F}+\mathcal{O}(\alpha^{'})
\label{Fds2}
\end{eqnarray}
where $B\in\Omega^1(M_8)$, $X\in\Omega^3(M_8)$,  $f\in\Omega^0(M_8)$ and $\mathcal{F}$ is a 2-form on $M_{8}$ taking values in $\mathfrak{g}$.

\end{itemize}

\subsection{Reduction of the gravitino KSE on $M_{10-n}$}

In this sub-section we reduce the gravitino KSE \eqref{KSEcorr1} on $M_{10-n}$, for generic $n$. Using \eqref{spinconnection2}, \eqref{Hdsn}, \eqref{Hds3} and \eqref{Hds2}, the $M=\mu$ component of \eqref{KSEcorr1} is given by
\begin{eqnarray}
\frac{\partial}{\partial x^{\mu}}\epsilon=\frac{1}{\mathcal{U}}\bigg(-\frac{k}{4}x^{\nu}\Gamma_{\nu\mu}+\Gamma_{\mu}\mathcal{C}_{(n)}\bigg )\epsilon+\mathcal{O}(\alpha^{'2})
\label{mudesitD}
\end{eqnarray}
where we have defined\footnote{If $\omega\in\Omega^p(M_{10-n})$, then $\slashed{\omega}=\omega_{a_1a_2\dots a_p}\Gamma^{a_1a_2\dots a_p}$.} 
\begin{eqnarray}
\mathcal{C}_{(n)}=
\begin{cases}
-\frac{1}{2}\widetilde{\slashed{\nabla}}A~~~~~~~~~~~~~~~~~~~~~~~n\ge 4 \\\\
-\frac{1}{2}\widetilde{\slashed{\nabla}}A+\frac{1}{4}cA^{-2}\Gamma^{(3)}~~~~~~\,\,n=3 \\\\
-\frac{1}{2}\widetilde{\slashed{\nabla}}A+\frac{1}{4}A^{-1}\slashed{B}\Gamma^{(2)}~~~~~~n=2
\end{cases}
\label{CN}
\end{eqnarray}
with $\Gamma^{(n)}$ being the highest rank Gamma matrix on $dS_n$, that is
\begin{eqnarray}
\Gamma^{(n)}:=\frac{1}{n!}\epsilon_{\mu_1\mu_2\dots \mu_n}\Gamma^{\mu_1\mu_2\dots \mu_n}~.
\end{eqnarray}
Equation \eqref{mudesitD} yields a partial differential equation for $\epsilon$
\begin{eqnarray}
\frac{\partial}{\partial x^{\mu}}\frac{\partial}{\partial x^{\nu}}\epsilon+\frac{k}{4\mathcal{U}}\big(x_{\mu}\frac{\partial}{\partial x^{\nu}}\epsilon+x_{\nu}\frac{\partial}{\partial x^{\mu}}\epsilon\big)-\frac{k^2}{16\mathcal{U}^2}x_{\mu}x_{\nu}\epsilon+\frac{k}{4\mathcal{U}}\eta_{\mu\nu}\epsilon=\mathcal{O}(\alpha^{'2})
\nonumber \\
\label{ksemu14}
\end{eqnarray}
whose solution is given by
\begin{eqnarray}
\epsilon=\mathcal{U}^{-1/2}\big(\psi+x^{\mu}\tau_{\mu}\big)+\mathcal{O}(\alpha^{'2})
\label{ksemu21}
\end{eqnarray}
where $\psi$ and $\tau_{\mu}$ are 16-component chiral spinors which depend only on the co-ordinates on $M_{10-n}$. Inserting \eqref{ksemu21} into \eqref{mudesitD}, we get
\begin{eqnarray}
\tau_{\mu}=\Gamma_{\mu}\mathcal{C}_{(n)}\psi+\mathcal{O}(\alpha^{'2})
\label{ksemu24}
\end{eqnarray}
and
\begin{eqnarray}
\bigg(\frac{k}{4}+\widehat{\mathcal{C}}_{(n)}\mathcal{C}_{(n)}\bigg)\psi=\mathcal{O}(\alpha^{'2})
\label{ksemu26}
\end{eqnarray}
where $\widehat{\mathcal{C}}_{(n)}$ is defined by
\begin{eqnarray}
\widehat{\mathcal{C}}_{(n)}=
\begin{cases}
\frac{1}{2}\widetilde{\slashed{\nabla}}A~~~~~~~~~~~~~~~~~~~~~~n\ge 4 \\\\
\frac{1}{2}\widetilde{\slashed{\nabla}}A+\frac{1}{4}cA^{-2}\Gamma^{(3)}~~~~~~n=3 \\\\
\frac{1}{2}\widetilde{\slashed{\nabla}}A+\frac{1}{4}A^{-1}\slashed{B}\Gamma^{(2)}~~~~~n=2
\end{cases}
\end{eqnarray}
Furthermore, substituting \eqref{ksemu24} into \eqref{ksemu21}, we find
\begin{eqnarray}
\epsilon=\mathcal{U}^{-1/2}\bigg(1+x^{\mu}\Gamma_{\mu}\mathcal{C}_{(n)}\bigg)\psi+\mathcal{O}(\alpha^{'2})~.
\label{epsilonfinale}
\end{eqnarray}
Moreover, the $M=a$ component of \eqref{KSEcorr1} reads 
\begin{eqnarray}
\widetilde{\nabla}_a\epsilon=\sigma_{a(n)}\epsilon+\mathcal{O}(\alpha^{'2})
\label{adsn}
\end{eqnarray}
where $\sigma_{a(n)}$ is given by
\begin{eqnarray}
\sigma_{a(n)}=
\begin{cases}
\frac{1}{8}X_{abc}\Gamma^{bc}~~~~~~~~~~~~~~~~~~~~~~~n \ge 3 \\\\
\frac{1}{8}X_{abc}\Gamma^{bc}+\frac{1}{4}A^{-2}B_a\Gamma^{(2)} ~~~~n=2
\end{cases}
\label{sigman}
\end{eqnarray}
Substituting \eqref{epsilonfinale} into \eqref{adsn}, we find
\begin{eqnarray}
\widetilde{\nabla}_a\psi=\sigma_{a(n)}\psi+\mathcal{O}(\alpha^{'2})
\label{ksem9}
\end{eqnarray}
and
\begin{eqnarray}
\bigg(\widetilde{\nabla}_a\mathcal{C}_{(n)}+\mathcal{C}_{(n)}\sigma_{a(n)}-\widehat{\sigma}_{a(n)}\mathcal{C}_{(n)}\bigg)\psi=\mathcal{O}(\alpha^{'2})
\label{ksem92}
\end{eqnarray}
where $\hat{\sigma}_{a(n)}$ is defined by
\begin{eqnarray}
\widehat{\sigma}_{a(n)}=
\begin{cases}
\frac{1}{8}X_{abc}\Gamma^{bc}~~~~~~~~~~~~~~~~~~~~~~~n \ge 3 \\\\
\frac{1}{8}X_{abc}\Gamma^{bc}-\frac{1}{4}A^{-2}B_a\Gamma^{(2)} ~~~~n=2
\end{cases}
\label{sigman}
\end{eqnarray}
Hence, the reduction of the gravitino KSEs \eqref{KSEcorr1} on $M_{10-n}$ yield the gravitino KSE \eqref{ksem9} on $M_{10-n}$, supplemented by the quadratic conditions \eqref{ksemu26} and \eqref{ksem92}.

\section{$dS_n$, with $n\ge 4$}

In this section, we study warped $dS_n$ backgrounds, with $n\ge 4$. Using \eqref{CN} and \eqref{epsilonfinale}, we find 
\begin{eqnarray}
\epsilon=\mathcal{U}^{-1/2}\bigg(1-\frac{1}{2}x^{\mu}\Gamma_{\mu}\widetilde{\slashed{\nabla}}A\bigg)\psi+\mathcal{O}(\alpha^{'2})
\label{epsilon}
\end{eqnarray}
where $\mathcal{U}$ is defined by \eqref{R}. Moreover, the reduction of the gravitino KSE \eqref{KSEcorr1} on $M_{10-n}$ yields \eqref{ksem9}, that is
\begin{eqnarray}
\widetilde{\nabla}_a\psi=\frac{1}{8}X_{abc}\Gamma^{bc}\psi+\mathcal{O}(\alpha^{'2})
\label{gravino}
\end{eqnarray}
supplemented by the quadratic conditions \eqref{ksemu26} and \eqref{ksem92}, which imply
\begin{eqnarray}
k-(\widetilde{\nabla}A)^2=\mathcal{O}(\alpha^{'2})
\label{useful87}
\end{eqnarray}
and
\begin{eqnarray}
\bigg(-\frac{1}{2}\widetilde{\nabla}_a\widetilde{\nabla}_bA+\frac{1}{4}\widetilde{\nabla}^cAX_{cab}\bigg)\Gamma^b\psi=\mathcal{O}(\alpha^{'2})
\label{aux1}
\end{eqnarray}
respectively. Equation \eqref{aux1} yields
\begin{eqnarray}
-\frac{1}{2}\widetilde{\nabla}_a\widetilde{\nabla}_bA+\frac{1}{4}\widetilde{\nabla}^cAX_{cab}=\mathcal{O}(\alpha^{'2})
\label{aux2}
\end{eqnarray}
Splitting \eqref{aux2} into symmetric and anti-symmetric part in $(a,b)$, we get
\begin{eqnarray}
\widetilde{\nabla}_a\widetilde{\nabla}_bA=\mathcal{O}(\alpha^{'2})
\label{useful88}
\end{eqnarray}
and
\begin{eqnarray}
\widetilde{\nabla}^cAX_{cab}=\mathcal{O}(\alpha^{'2})~.
\label{useful89}
\end{eqnarray}
Furthermore, using \eqref{epsilon}, \eqref{Hdsn} and \eqref{Fdsn}, the dilatino and the gaugino KSE \eqref{KSEcorr2} and \eqref{KSEcorr3} decompose on $M_{10-n}$ as
\begin{eqnarray}
\bigg(\widetilde{\slashed{\nabla}}\Phi-\frac{1}{12}\slashed{X}\bigg)\psi=\mathcal{O}(\alpha^{'2})
\label{dilatcor}
\end{eqnarray}
\begin{eqnarray}
\slashed{\mathcal{F}}\psi=\mathcal{O}(\alpha^{'})
\label{gauginocor}
\end{eqnarray}
supplemented by the quadratic conditions
\begin{eqnarray}
\bigg(\widetilde{\slashed{\nabla}}\Phi-\frac{1}{12}\slashed{X}\bigg)\widetilde{\slashed{\nabla}}A\psi=\mathcal{O}(\alpha^{'2})
\label{quadcor3}
\end{eqnarray}
\begin{eqnarray}
\slashed{\mathcal{F}}\widetilde{\slashed{\nabla}}A\psi=\mathcal{O}(\alpha^{'})~.
\label{quadcor4}
\end{eqnarray}
Combining \eqref{dilatcor} and \eqref{quadcor3} and using \eqref{useful89}, we obtain
\begin{eqnarray}
\widetilde{\nabla}^aA\widetilde{\nabla}_a\Phi=\mathcal{O}(\alpha^{'2})~.
\label{useful90}
\end{eqnarray}
Moreover, combining \eqref{gauginocor} and \eqref{quadcor4}, we get
\begin{eqnarray}
\widetilde{\nabla}^aA \mathcal{F}_{ab}=\mathcal{O}(\alpha^{'})~.
\label{useful91}
\end{eqnarray}
Using \eqref{useful87}, \eqref{useful88} and \eqref{useful89}, a straightforward computation shows that the only non-vanishing components of $\check{R}$ up to two loops are given by 
\begin{eqnarray}
\tilde{\check{R}}_{ab,cd}:=\widetilde{R}_{abcd}-\frac{1}{2}\widetilde{\nabla}_a X_{cbd}+\frac{1}{2}\widetilde{\nabla}_b X_{cad}+\frac{1}{4}X_{cam}X^m_{~~bd}-\frac{1}{4}X_{cbm}X^m_{~~ad}~.
\nonumber \\
\label{tildecheckR}
\end{eqnarray}
In order to obtain \eqref{tildecheckR}, we have used \eqref{checkR}, \eqref{spinconnection2} and \eqref{Riemann}. Using the previous results, the bosonic field equations \eqref{Einsteincorr}-\eqref{dilatoncorr} decompose on $M_{10-n}$ as follows
\begin{eqnarray}
\widetilde{R}_{ab}-\frac{1}{4}X_{amn}X_b^{~mn}+2\widetilde{\nabla}_a\widetilde{\nabla}_b\Phi+\frac{\alpha^{'}}{4}\bigg(\tilde{\check{R}}_{ac_1,c_2c_3}\tilde{\check{R}}_{b}^{~~c_1,c_2c_3}-\mathcal{F}_{ac ij}\mathcal{F}_b^{~cij}\bigg)=\mathcal{O}(\alpha^{'2})
\nonumber \\
\label{bfied111}
\end{eqnarray}
\begin{eqnarray}
(\widetilde{\nabla}^cX)_{cab}-2\widetilde{\nabla}^c\Phi X_{cab}=\mathcal{O}(\alpha^{'2})
\end{eqnarray}
\begin{eqnarray}
\widetilde{\nabla}^b\mathcal{F}_{ba}-2\widetilde{\nabla}^b\Phi\mathcal{F}_{ba}+\frac{1}{2}X_{ab_1b_2}\mathcal{F}^{b_1b_2}=\mathcal{O}(\alpha^{'})
\end{eqnarray}
\begin{eqnarray}
\widetilde{\nabla}^a\widetilde{\nabla}_a\Phi-2(\widetilde{\nabla}\Phi)^2+\frac{1}{12}X^2-\frac{\alpha^{'}}{16}\bigg(\tilde{\check{R}}_{a_1a_2,a_3a_4}\tilde{\check{R}}^{a_1a_2,a_3a_4}-\mathcal{F}_{ab,ij}\mathcal{F}^{ab,ij}\bigg)=\mathcal{O}(\alpha^{'2})
\nonumber \\
\label{bfied222}
\end{eqnarray}
where we have used \eqref{Hdsn}, \eqref{Fdsn}, \eqref{spinconnection2} and \eqref{Ricci}. Moreover, the Bianchi identities \eqref{BianchiDH} imply
\begin{eqnarray}
\tilde{d}X=-\frac{\alpha^{'}}{4}\bigg(\textrm{tr}(\tilde{\check{R}}\wedge \tilde{\check{R}})-\textrm{tr}(\mathcal{F}\wedge\mathcal{F})\bigg)+\mathcal{O}(\alpha^{'2})~.
\label{bfied3}
\end{eqnarray}
Let us introduce the 1-form on $M_{10-n}$
\begin{eqnarray}
V=\frac{1}{\sqrt{k}}dA~.
\label{V1form}
\end{eqnarray}
Equation \eqref{useful88} implies
\begin{eqnarray}
\widetilde{\nabla}_a V_b=\mathcal{O}(\alpha^{'2})~.
\end{eqnarray}
Moreover, \eqref{useful87} implies $V^2=1+\mathcal{O}(\alpha^{'2})$. Hence
\begin{eqnarray}
ds^2(M_{10-n})=V\otimes V+ds^2(M_{9-n})+\mathcal{O}(\alpha^{'2})=\frac{1}{k}dA^2+ds^2(M_{9-n})+\mathcal{O}(\alpha^{'2})~.
\nonumber \\
\label{split1}
\end{eqnarray}
Substituting \eqref{split1} in \eqref{EQ1}, we find
\begin{eqnarray}
ds^2(M_{10})=ds^2(M_{1,n})+ds^2(M_{9-n})+\mathcal{O}(\alpha^{'2})
\label{auxm10}
\end{eqnarray}
where we have defined
\begin{eqnarray}
ds^2(M_{1,n})=A^2ds^2(dS_n)+\frac{1}{k}dA^2~.
\label{dsm1n}
\end{eqnarray}
It is straightforward to check that the Riemann tensor of \eqref{dsm1n} vanishes, hence $M_{1,n}\simeq \mathbb{R}^{1,n}$ and \eqref{auxm10} yields
\begin{eqnarray}
ds^2(M_{10})=ds^2(\mathbb{R}^{1,n})+ds^2(M_{9-n})+\mathcal{O}(\alpha^{'2})~.
\label{finaledec}
\end{eqnarray}
Notice that \eqref{useful89}, \eqref{useful90} and \eqref{useful91} imply that $X$, $\Phi$ and $\mathcal{F}$ have non-vanishing components only along $M_{9-n}$. Moreover, the vector field $V$ dual to the 1-form \eqref{V1form}\footnote{We denote the vector field dual to the 1-form \eqref{V1form} by the same symbol $V$.} preserves $X$, $\Phi$ and $\mathcal{F}$, that is
\begin{eqnarray}
\widetilde{\mathcal{L}}_V \Phi=\mathcal{O}(\alpha^{'2})
\label{LVphi}
\end{eqnarray}
\begin{eqnarray}
\widetilde{\mathcal{L}}_V \mathcal{F}=\mathcal{O}(\alpha^{'})
\label{LVFF}
\end{eqnarray}
\begin{eqnarray}
\widetilde{\mathcal{L}}_VX=\mathcal{O}(\alpha^{'2})~.
\label{nicex}
\end{eqnarray}
Equation \eqref{LVphi} is a consequence of \eqref{useful90}. Equation \eqref{LVFF} follows from \eqref{useful91}, by fixing the gauge $i_V\mathcal{A}=0$ where $\mathcal{A}$ is the gauge potential $\mathcal{A}$, defined by $\mathcal{F}=d\mathcal{A}+\mathcal{A}\wedge \mathcal{A}$.
In order to show \eqref{nicex}, first of all notice that
\begin{eqnarray}
\mathcal{L}_VX=i_V \tilde{d}X+\mathcal{O}(\alpha^{'2})
\label{ivdx}
\end{eqnarray}
by means of \eqref{useful89}. Moreover, implementing \eqref{epsilon} in the 10-dimensional gravitino integrability conditions
\begin{eqnarray}
\hat{R}_{MN,PQ}\Gamma^{PQ}\epsilon=\mathcal{O}(\alpha^{'2})
\end{eqnarray}
we find
\begin{eqnarray}
\hat{R}_{MN,PQ}\Gamma^{PQ}\psi=\mathcal{O}(\alpha^{'2})
\label{in56}
\end{eqnarray}
and
\begin{eqnarray}
\hat{R}_{MN,PQ}\Gamma^{PQ}\slashed{V}\psi=\mathcal{O}(\alpha^{'2})~.
\label{in57}
\end{eqnarray}
Taking $M,N$ to be internal indices in \eqref{in56} and \eqref{in57}, we get
\begin{eqnarray}
\tilde{\hat{R}}_{mn,ab}\Gamma^{ab}\psi=\mathcal{O}(\alpha^{'2})
\label{int54}
\end{eqnarray}
and
\begin{eqnarray}
\tilde{\hat{R}}_{mn,ab}\Gamma^{ab}\slashed{V}\psi=\mathcal{O}(\alpha^{'2})~.
\label{int55}
\end{eqnarray}
In order to obtain \eqref{int54} and \eqref{int55}, we have used $\hat{R}_{mn,\mu\nu}=\hat{R}_{mn,\mu a}=\mathcal{O}(\alpha^{'2})$ which follow from \eqref{hatR}. Equations \eqref{int54} and \eqref{int55} yield
\begin{eqnarray}
V^a\tilde{\hat{R}}_{mn,ab}=\mathcal{O}(\alpha^{'2})
\end{eqnarray}
hence, using \eqref{identity}, it follows that
\begin{eqnarray}
V^a \tilde{\check{R}}_{ab,mn}=\mathcal{O}(\alpha^{'})~.
\label{use78}
\end{eqnarray}
Implementing \eqref{useful91} and \eqref{use78} in the Bianchi identities \eqref{bfied3}, we find
\begin{eqnarray}
i_V \tilde{d}X=\mathcal{O}(\alpha^{'2})~.
\label{nicex2}
\end{eqnarray}
Substituting \eqref{nicex2} in \eqref{ivdx}, we recover \eqref{nicex}. To summarize, in heterotic perturbation theory, up to two loops, warped $dS_n$ backgrounds with $n\ge 4$ are direct product manifolds of $(n+1)$-Minkowski spacetime and a Riemannian manifold $M_{9-n}$. Moreover, the 1-form $d\Phi$, the 3-form $H$ and the non abelian field strength $F$ have non vanishing components only along $M_{9-n}$. Such backgrounds have been extensively studied in the literature, see e.g. \cite{Ivanov}, \cite{Ivanov2} and references within.

\section{$dS_3$ backgrounds}

In this section, we analyze warped $dS_3$ backgrounds, up to two loops. In this case, equation \eqref{ksemu26} yields
\begin{eqnarray}
\bigg(\frac{k}{4}-\frac{1}{4}(\widetilde{\nabla}A)^2+\frac{1}{4}fA^{-2}\widetilde{\slashed{\nabla}}A\Gamma^{(3)}+\frac{1}{16}f^2A^{-4}\bigg)\psi=\mathcal{O}(\alpha^{'2})~.
\label{intintint1}
\end{eqnarray}
Since $\Gamma^a\Gamma^{(3)}$ is anti-hermitian, equation \eqref{intintint1} implies
\begin{eqnarray}
\frac{k}{4}-\frac{1}{4}(\widetilde{\nabla}A)^2+\frac{1}{16}f^2A^{-4}=\mathcal{O}(\alpha^{'2})
\label{ccccc}
\end{eqnarray}
and
\begin{eqnarray}
f\widetilde{\slashed{\nabla}}A\Gamma^{(3)}\psi=\mathcal{O}(\alpha^{'2})~.
\label{dddd}
\end{eqnarray}
Equation \eqref{dddd} yields
\begin{eqnarray}
f\tilde{d}A=\mathcal{O}(\alpha^{'2})~.
\label{ff}
\end{eqnarray}
Moreover, the zero-th order term in $\alpha^{'}$ in the perturbation series of $\tilde{d}A$ is not vanishing, as a consequence of \eqref{ccccc}. Hence \eqref{ff} yields
\begin{eqnarray}
f=\mathcal{O}(\alpha^{'2})
\end{eqnarray}
that is $H=X+\mathcal{O}(\alpha^{'2})$ by means of \eqref{Hds3}, which in turn implies that the case $n=3$ can be treated exactly as $n\ge 4$. In particular, equation \eqref{finaledec} still holds for $n=3$, that is
\begin{eqnarray}
ds^2(M_{10})=ds^2(\mathbb{R}^{1,3})+ds^2(M_6)+\mathcal{O}(\alpha^{'2})~.
\end{eqnarray}
Moreover, equations \eqref{useful89}, \eqref{useful90} and \eqref{useful91} still hold, i.e.
\begin{eqnarray}
i_V X=\mathcal{O}(\alpha^{'2})~,~~~i_Vd\Phi=\mathcal{O}(\alpha^{'2})~,~~~i_V \mathcal{F}=\mathcal{O}(\alpha^{'})~.
\end{eqnarray}
To summarize, in heterotic perturbation theory, up to two loops, warped $dS_3$ backgrounds are $\mathbb{R}^{1,3}\times M_6$. Moreover, the 1-form $d\Phi$, the 3-form $H$ and the non abelian field strength $F$ have non vanishing components only along $M_{6}$.

\section{$dS_2$}

In this section, we study warped product $dS_2$ backgrounds, up to two loops. First of all, equation \eqref{epsilonfinale} reads
\begin{eqnarray}
\epsilon=\mathcal{U}^{-1/2}\bigg(1+x^{\mu}\Gamma_{\mu}\big(-\frac{1}{2}\widetilde{\slashed{\nabla}}A+\frac{1}{4}A^{-1}\slashed{B}\Gamma^{(2)}\big)\bigg)\psi+\mathcal{O}(\alpha^{'2})~.
\label{epsilonds2}
\end{eqnarray}
Moreover, the reduction of the gravitino KSE \eqref{KSEcorr1} on $M_{10-n}$ yields \eqref{ksem9}, that is
\begin{eqnarray}
\widetilde{\nabla}_a\psi=\bigg(\frac{1}{8}X_{abc}\Gamma^{bc}+\frac{1}{4}A^{-2}B_a\Gamma^{(2)}\bigg)\psi+\mathcal{O}(\alpha^{'2})
\label{gravitcords2}
\end{eqnarray}
supplemented by the quadratic constraints \eqref{ksemu26} and \eqref{ksem92}, which are equivalent to the conditions
\begin{eqnarray}
\bigg(\frac{k}{4}-\frac{1}{4}(\widetilde{\nabla}A)^2+\frac{1}{4}A^{-1}\widetilde{\nabla}_a AB_b\Gamma^{ab}\Gamma^{(2)}+\frac{1}{16}A^{-2}B^2\bigg)\psi=\mathcal{O}(\alpha^{'2})
\label{quadcor75}
\end{eqnarray}
and
\begin{eqnarray}
&&\bigg(-\frac{1}{2}\Gamma^b\widetilde{\nabla}_a\widetilde{\nabla}_b A-\frac{1}{4}A^{-2}\widetilde{\nabla}_a A B_b\Gamma^b\Gamma^{(2)}+\frac{1}{4}A^{-1}\widetilde{\nabla}_a B_b\Gamma^b\Gamma^{(2)}+\frac{1}{4}\widetilde{\nabla}^dAX_{dab}\Gamma^b
\nonumber \\
&&-\frac{1}{4}A^{-2}B_a\widetilde{\nabla}_b A\Gamma^b\Gamma^{(2)}-\frac{1}{8}A^{-1}B^dX_{dab}\Gamma^b\Gamma^{(2)}+\frac{1}{8}A^{-3}B_aB_b\Gamma^b\bigg)\psi=\mathcal{O}(\alpha^{'2})
\nonumber \\
\label{quadcor76}
\end{eqnarray}
respectively. Furthermore, using \eqref{epsilonds2}, \eqref{Hds2} and \eqref{Fds2}, the dilatino and the gaugino KSE \eqref{KSEcorr2} and \eqref{KSEcorr3} decompose on $M_{10-n}$ as
\begin{eqnarray}
\bigg(\widetilde{\slashed{\nabla}}\Phi-\frac{1}{12}\slashed{X}-\frac{1}{2}A^{-2}\slashed{B}\Gamma^{(2)}\bigg)\psi=\mathcal{O}(\alpha^{'2})
\label{dilatcords2}
\end{eqnarray}
and 
\begin{eqnarray}
2fA^{-2}\Gamma^{(2)}\psi+\slashed{\mathcal{F}}\psi=\mathcal{O}(\alpha^{'})~.
\label{gauginocords2}
\end{eqnarray}
Notice that equation \eqref{gauginocords2} is equivalent to
\begin{eqnarray}
2fA^{-2}\psi+\Gamma^{(2)}\slashed{\mathcal{F}}\psi=\mathcal{O}(\alpha^{'})~.
\label{simplif}
\end{eqnarray}
Equation \eqref{simplif} implies $f=\mathcal{O}(\alpha^{'})$, that is $F=\mathcal{F}+\mathcal{O}(\alpha^{'})$ and
\begin{eqnarray}
\slashed{\mathcal{F}}\psi=\mathcal{O}(\alpha^{'})~.
\label{gauginofinal}
\end{eqnarray}
Moreover, we find the quadratic constraints
\begin{eqnarray}
\bigg(\widetilde{\slashed{\nabla}}\Phi-\frac{1}{12}\slashed{X}+\frac{1}{2}A^{-2}\slashed{B}\Gamma^{(2)}\bigg)\bigg(-\frac{1}{2}\widetilde{\slashed{\nabla}}A+\frac{1}{4}A^{-1}\slashed{B}\Gamma^{(2)}\bigg)\psi=\mathcal{O}(\alpha^{'2})
\label{quadcor77}
\end{eqnarray}
\begin{eqnarray}
\slashed{\mathcal{F}}\bigg(-\frac{1}{2}\widetilde{\slashed{\nabla}}A+\frac{1}{4}A^{-1}\slashed{B}\Gamma^{(2)}\bigg)\psi=\mathcal{O}(\alpha^{'})~.
\label{quadcor78}
\end{eqnarray}
Equation \eqref{quadcor75} yields
\begin{eqnarray}
\frac{k}{4}-\frac{1}{4}(\widetilde{\nabla}A)^2+\frac{1}{16}A^{-2}B^2=\mathcal{O}(\alpha^{'2})
\label{useful1}
\end{eqnarray}
and
\begin{eqnarray}
\widetilde{\nabla}_a AB_b\Gamma^{ab}\psi=\mathcal{O}(\alpha^{'2}).
\label{useful2}
\end{eqnarray}
Equation \eqref{useful1} implies that $dA$ is not vanishing, at zero-th order in $\alpha^{'}$. Hence, point-wise, we can choose a frame such that, at zero-th order in $\alpha^{'}$, $\widetilde{\nabla}_2 A\ne 0$, $\widetilde{\nabla}_3 A=0,\dots, \widetilde{\nabla}_9 A=0$, where we have denoted by $2,3,\dots 9$, the directions along $M_8$. Thus \eqref{useful2} yields
\begin{eqnarray}
B_i \Gamma^i\psi=\mathcal{O}(\alpha^{'2})
\label{us54}
\end{eqnarray}
where $i=3,4,\dots 9$. Equation \eqref{us54} yields $B_i=\mathcal{O}(\alpha^{'2})$, which means that 
\begin{eqnarray}
B=\mathcal{G} \tilde{d}A+\mathcal{O}(\alpha^{'2})~.
\label{BfDA}
\end{eqnarray}
for some function $\mathcal{G}$. The integrability conditions of the 10-dimensional gravitino KSE \eqref{KSEcorr1} read
\begin{eqnarray}
\hat{R}_{MN,RS}\Gamma^{RS}\epsilon=\mathcal{O}(\alpha^{'2})~.
\label{integrab10}
\end{eqnarray}
Taking $M,N$ to be internal indices in \eqref{integrab10}, we obtain
\begin{eqnarray}
\hat{R}_{ab,\mu\nu}\Gamma^{\mu\nu}\epsilon+2\hat{R}_{ab,\nu c}\Gamma^{\nu c}\epsilon+\hat{R}_{ab,cd}\Gamma^{cd}\epsilon=\mathcal{O}(\alpha^{'2})~.
\label{integrab101}
\end{eqnarray}
Using \eqref{hatR}, a direct computation yields $\hat{R}_{ab,\nu c}=\mathcal{O}(\alpha^{'2})$, hence \eqref{integrab101} reads
\begin{eqnarray}
\hat{R}_{ab,\mu\nu}\epsilon^{\mu\nu}\epsilon +\hat{R}_{ab,cd}\Gamma^{(2)}\Gamma^{cd}\epsilon=\mathcal{O}(\alpha^{'2})~.
\label{integrab102}
\end{eqnarray}
Equation \eqref{integrab102} implies 
\begin{eqnarray}
\hat{R}_{ab,\mu\nu}=\mathcal{O}(\alpha^{'2})
\label{integrab103}
\end{eqnarray}
which is equivalent to
\begin{eqnarray}
-\frac{1}{2}A^{-2}(\tilde{d}B)_{ab}+A^{-3}(dA\wedge B)_{ab}=\mathcal{O}(\alpha^{'2})~.
\label{dBB}
\end{eqnarray}
Using \eqref{BfDA}, we have that $\tilde{d}A\wedge B=\mathcal{O}(\alpha^{'2})$, hence \eqref{dBB} yields
\begin{eqnarray}
\tilde{d}B=\mathcal{O}(\alpha^{'2})
\label{bianchicorr}
\end{eqnarray}
hence $\mathcal{G}=\mathcal{G}(A)+\mathcal{O}(\alpha^{'2})$. Notice that if $\psi$ is Killing, i.e. if it satisfies \eqref{gravitcords2}, \eqref{dilatcords2}, \eqref{gauginofinal}, \eqref{quadcor75}, \eqref{quadcor76}, \eqref{quadcor77} and \eqref{quadcor78}, then $\Gamma^{(2)}\psi$ is Killing, since $[\Gamma^{(2)},\Gamma^a]=0$. Moreover $(\Gamma^{(2)})=1$, hence w.l.o.g. we can set
\begin{eqnarray}
\Gamma^{(2)}\psi=\psi~.
\label{projproj}
\end{eqnarray}
Substituting \eqref{projproj} and \eqref{BfDA} in \eqref{quadcor76}, we find
\begin{eqnarray}
&&-\frac{1}{2}\widetilde{\nabla}_a\widetilde{\nabla}_bA-\frac{1}{2}A^{-2}\mathcal{G}\widetilde{\nabla}_a A\widetilde{\nabla}_b A+\frac{1}{4}A^{-1}\widetilde{\nabla}_a B_b+\frac{1}{4}\widetilde{\nabla}^d AX_{dab}
\nonumber \\
&&-\frac{1}{8}A^{-1}\mathcal{G}\widetilde{\nabla}^dAX_{dab}+\frac{1}{8}A^{-3}\mathcal{G}^2\widetilde{\nabla}_a A\widetilde{\nabla}_b A=\mathcal{O}(\alpha^{'2})~.
\label{conduseful}
\end{eqnarray}
In the following, we define the function $g$ by $\mathcal{G}:=A^2g$, that is 
\begin{eqnarray}
B=A^2g dA+\mathcal{O}(\alpha^{'2})~.
\label{otherus3}
\end{eqnarray}
Notice that $g=g(A)+\mathcal{O}(\alpha^{'2})$ by means of \eqref{bianchicorr}. Using \eqref{bianchicorr}, the anti-symmetric part of \eqref{conduseful} yields
\begin{eqnarray}
\bigg(1-\frac{1}{2}Ag\bigg)\widetilde{\nabla}^dAX_{dab}=\mathcal{O}(\alpha^{'2})~.
\label{otherus2}
\end{eqnarray}
Notice that \eqref{useful1} implies
\begin{eqnarray}
\frac{k}{4}-\frac{1}{4}\bigg(1-\frac{1}{4}A^2g^2\bigg)(\widetilde{\nabla}A)^2=\mathcal{O}(\alpha^{'2})
\label{otherus}
\end{eqnarray}
If $1-\frac{1}{2}Ag=\mathcal{O}(\alpha^{'})$, then $(1+\frac{1}{2}Ag)(1-\frac{1}{2}Ag)=1-\frac{1}{4}A^2g^2=\mathcal{O}(\alpha^{'})$, which is in contradiction with respect to \eqref{otherus}. Hence $1-\frac{1}{2}A g$ has non-vanishing zero-th term in $\alpha^{'}$ and \eqref{otherus2} yields
\begin{eqnarray}
\widetilde{\nabla}^dAX_{dab}=\mathcal{O}(\alpha^{'2})~.
\label{daX}
\end{eqnarray}
Moreover, inserting \eqref{otherus3} in \eqref{conduseful}, the symmetric part of \eqref{conduseful} yields
\begin{eqnarray}
-\frac{1}{2}\widetilde{\nabla}_a\widetilde{\nabla}_bA+\frac{1}{4}A\widetilde{\nabla}_a (g\widetilde{\nabla}_b A)+\frac{1}{8}Ag^2\widetilde{\nabla}_aA\widetilde{\nabla}_bA=\mathcal{O}(\alpha^{'2})~.
\label{symuseful}
\end{eqnarray}
We will come back to \eqref{symuseful} later. Next, notice that equations \eqref{otherus3} and \eqref{projproj} imply
\begin{eqnarray}
\bigg(-\frac{1}{2}\widetilde{\slashed{\nabla}}A+\frac{1}{4}A^{-1}\slashed{B}\Gamma^{(2)}\bigg)\psi=-\frac{1}{2}\big(1-\frac{1}{2}Ag\big)\widetilde{\slashed{\nabla}}A\psi
\end{eqnarray}
hence \eqref{quadcor77} and \eqref{quadcor78} can be rewritten as
\begin{eqnarray}
\bigg(\widetilde{\slashed{\nabla}}\Phi-\frac{1}{12}\slashed{X}+\frac{1}{2}A^{-2}\slashed{B}\bigg)\widetilde{\slashed{\nabla}}A\psi=\mathcal{O}(\alpha^{'2})
\label{quadcorr77bis}
\end{eqnarray}
and
\begin{eqnarray}
\slashed{\mathcal{F}}\widetilde{\slashed{\nabla}}A\psi=\mathcal{O}(\alpha^{'})~.
\label{quadcorr78bis}
\end{eqnarray}
Using \eqref{gauginofinal} and \eqref{quadcorr78bis}, we find
\begin{eqnarray}
\widetilde{\nabla}^b A\mathcal{F}_{ab}=\mathcal{O}(\alpha^{'})~.
\label{daF}
\end{eqnarray}
Furthermore, utilizing \eqref{dilatcords2} and \eqref{quadcorr77bis}, we find
\begin{eqnarray}
\widetilde{\nabla}^a A\widetilde{\nabla}_a\Phi=\mathcal{O}(\alpha^{'2})
\label{DAPHI}
\end{eqnarray}
where we have used \eqref{daX}. Implementing \eqref{useful1}, \eqref{bianchicorr}, \eqref{otherus3}, \eqref{daX} and \eqref{symuseful} in \eqref{checkR}, we find that the non-vanishing components of $\check{R}$ are given by
\begin{eqnarray}
\check{R}_{\mu a, b\rho}=\frac{1}{2}\big(\eta_{\mu\rho}+\epsilon_{\mu\rho}\big)\widetilde{\nabla}_a (g\widetilde{\nabla}_b A)
\label{checkrr}
\end{eqnarray}
and
\begin{eqnarray}
\check{R}_{ab,cd}=\tilde{\check{R}}_{ab,cd}~.
\label{checkrr2}
\end{eqnarray}
Now consider the $\alpha^{'}$ terms in the $(\mu,\nu)$ component of the Einstein equations \eqref{Einsteincorr}. First of all
\begin{eqnarray}
F_{\mu L ij}F_{\nu}^{~L ij}=\mathcal{O}(\alpha^{'2})~.
\label{alphaprime1}
\end{eqnarray}
Moreover, using \eqref{checkrr}, compute\footnote{On $dS_2$ $\epsilon_{\mu\sigma}\epsilon_{\nu}^{~\sigma}=-\eta_{\mu\nu}$.}
\begin{eqnarray}
\check{R}_{\mu L_1, L_2L_3}\check{R}_{\nu}^{~L_1,L_2L_3}&=&2\check{R}_{\mu a, \sigma b}\check{R}_{\nu}^{~a,\sigma b}
\nonumber \\
&=&(\eta_{\mu\sigma}+\epsilon_{\mu\sigma})(\delta_{\nu}^{~\sigma}+\epsilon_{\nu}^{~\sigma})C_{ab}C^{ab}+\mathcal{O}(\alpha^{'2})
\nonumber \\
&=&\mathcal{O}(\alpha^{'2})
\label{alphaprime2}
\end{eqnarray}
where we have defined $C_{ab}:=\frac{1}{2}\widetilde{\nabla}_a(g \widetilde{\nabla}_b A)$. Using \eqref{alphaprime1} and \eqref{alphaprime2}, the $(\mu,\nu)$ component of the Einstein equations \eqref{Einsteincorr} reads
\begin{eqnarray}
kA^{-2}-A^{-1}\widetilde{\nabla}^a\widetilde{\nabla}_a A-A^{-2}(\widetilde{\nabla}A)^2+\frac{1}{2}g^2(\widetilde{\nabla}A)^2=\mathcal{O}(\alpha^{'2})
\label{eino2}
\end{eqnarray}
where we have used \eqref{DAPHI}. Moreover, the gauge field equations \eqref{gaugeHcorr} yield
\begin{eqnarray}
\widetilde{\nabla}^aB_a-2A^{-1}\widetilde{\nabla}^a AB_a=\mathcal{O}(\alpha^{'2})
\label{aux99}
\end{eqnarray}
by implementing \eqref{DAPHI}. Inserting \eqref{otherus3} in \eqref{aux99}, we have
\begin{eqnarray}
A^2\frac{dg}{dA}(\widetilde{\nabla}A)^2+A^2 g\widetilde{\nabla}^a\widetilde{\nabla}_a A=\mathcal{O}(\alpha^{'2})~.
\label{gaugegood}
\end{eqnarray}
Combining \eqref{eino2} and \eqref{gaugegood} with \eqref{otherus}, we find that $g(A)$ satisfies the ODE
\begin{eqnarray}
\frac{dg}{dA}+\frac{1}{4}g^3 A=\mathcal{O}(\alpha^{'2})
\label{dgda}
\end{eqnarray}
whose solution is given by
\begin{eqnarray}
g=\frac{\eta}{\sqrt{q+\frac{A^2}{4}}}+\mathcal{O}(\alpha^{'2})
\label{gdef}
\end{eqnarray}
where $\eta^2=1$ and $q$ is a constant up to two loops. Notice that $q$ is not vanishing, as a consequence of \eqref{otherus}. Inserting \eqref{dgda} into \eqref{symuseful}, we find
\begin{eqnarray}
\widetilde{\nabla}_a W_b=\mathcal{O}(\alpha^{'2})
\label{nablaacons}
\end{eqnarray}
where we have defined $W:=g dA$. Notice that \eqref{nablaacons} and \eqref{checkrr} imply that the only non-vanishing component of $\check{R}$ up to two loops is \eqref{checkrr2}. Moreover, using \eqref{gdef} and \eqref{otherus}, we find
\begin{eqnarray}
W^2=\frac{k}{q}+\mathcal{O}(\alpha^{'2})~.
\label{s2}
\end{eqnarray}
Equation \eqref{s2} implies that $q$ is positive. Inserting \eqref{nablaacons} in \eqref{symuseful}, we find
\begin{eqnarray}
-2A^{-1}\widetilde{\nabla}_a\widetilde{\nabla}_bA+\frac{1}{2}g^2\widetilde{\nabla}_a\widetilde{\nabla}_b A=\mathcal{O}(\alpha^{'2})
\label{condus}
\end{eqnarray}
Using \eqref{condus}, the $(a,b)$-component of the Einstein equations \eqref{Einsteincorr} read
\begin{eqnarray}
\widetilde{R}_{ab}-\frac{1}{4}X_{amn}X_b^{~mn}+2\widetilde{\nabla}_a\widetilde{\nabla}_b\Phi+\frac{\alpha^{'}}{4}\bigg(\tilde{\check{R}}_{ac_1,c_2c_3}\tilde{\check{R}}_{b}^{~~c_1,c_2c_3}-\mathcal{F}_{ac ij}\mathcal{F}_b^{~cij}\bigg)=\mathcal{O}(\alpha^{'2})
\nonumber \\
\label{bfied11}
\end{eqnarray}
Moreover, the $(a,b)$-component of the gauge field equations \eqref{gaugeHcorr} yields
\begin{eqnarray}
(\widetilde{\nabla}^cX)_{cab}-2\widetilde{\nabla}^c\Phi X_{cab}=\mathcal{O}(\alpha^{'2})
\end{eqnarray}
The non-trivial component of \eqref{gaugeFcorr} yields
\begin{eqnarray}
\widetilde{\nabla}^b\mathcal{F}_{ba}-2\widetilde{\nabla}^b\Phi\mathcal{F}_{ba}+\frac{1}{2}X_{ab_1b_2}\mathcal{F}^{b_1b_2}=\mathcal{O}(\alpha^{'})
\end{eqnarray}
and the dilaton equation \eqref{dilatoncorr} is
\begin{eqnarray}
\widetilde{\nabla}^a\widetilde{\nabla}_a\Phi-2(\widetilde{\nabla}\Phi)^2+\frac{1}{12}X^2-\frac{\alpha^{'}}{16}\bigg(\tilde{\check{R}}_{a_1a_2,a_3a_4}\tilde{\check{R}}^{a_1a_2,a_3a_4}-\mathcal{F}_{ab,ij}\mathcal{F}^{ab,ij}\bigg)=\mathcal{O}(\alpha^{'2})~.
\nonumber \\
\label{bfied22}
\end{eqnarray}
Moreover, the Bianchi identities \eqref{BianchiDH} read
\begin{eqnarray}
\tilde{d}X=-\frac{\alpha^{'}}{4}\bigg(\textrm{tr}(\tilde{\check{R}}\wedge \tilde{\check{R}})-\textrm{tr}(\mathcal{F}\wedge\mathcal{F})\bigg)+\mathcal{O}(\alpha^{'2})~.
\label{bfied33}
\end{eqnarray}
Equation \eqref{nablaacons} implies that
\begin{eqnarray}
ds^2(M_8)&=&\frac{1}{W^2}W\otimes W+ds^2(M_7)+\mathcal{O}(\alpha^{'2})
\nonumber \\
&=&\frac{q}{k}\frac{dA^2}{q+\frac{A^2}{4}}+ds^2(M_7)+\mathcal{O}(\alpha^{'2})~.
\label{dsm8}
\end{eqnarray}
Substituting \eqref{dsm8} in \eqref{EQ1}, we get
\begin{eqnarray}
ds^2(M_{10})=ds^2(M_3)+ds^2(M_7)+\mathcal{O}(\alpha^{'2})
\label{auxm10ds2}
\end{eqnarray}
where
\begin{eqnarray}
ds^2(M_3)=A^2\frac{\eta_{\mu\nu}dx^{\mu}dx^{\nu}}{\big(1+\frac{k}{4}|x|^2\big)^2}+\frac{q}{k}\frac{dA^2}{q+\frac{A^2}{4}}+\mathcal{O}(\alpha^{'2})~.
\label{M3metric}
\end{eqnarray}
Define $y=\sqrt{k}x$, $A=\sqrt{q}\tilde{A}$ and $c^2=\frac{q}{k}$. Then \eqref{M3metric} can be rewritten solely in terms of $c^2$ as follows
\begin{eqnarray}
ds^2(M_3)=c^2\tilde{A}^2\frac{\eta_{\mu\nu}dy^{\mu}dy^{\nu}}{\big(1+\frac{1}{4}|y|^2\big)^2}+c^2\frac{d\tilde{A}^2}{1+\frac{\tilde{A}^2}{4}}+\mathcal{O}(\alpha^{'2})~.
\label{M3metric2}
\end{eqnarray}
The Riemann tensor of \eqref{M3metric2} is
\begin{eqnarray}
R_{\alpha\beta\gamma\delta}(M_3)=-\frac{1}{4c^2}\big(g_{\alpha\gamma}g_{\beta\delta}-g_{\alpha\delta}g_{\beta\gamma}\big)+\mathcal{O}(\alpha^{'2})
\end{eqnarray}
where $\alpha\in\{y^{\mu},\tilde{A}\}$. Hence $M_3\simeq AdS_3 +\mathcal{O}(\alpha^{'2})$ and \eqref{auxm10ds2} yields
\begin{eqnarray}
ds^2(M_{10})=ds^2(AdS_3)+ds^2(M_7)+\mathcal{O}(\alpha^{'2})
\end{eqnarray}
Moreover, equations \eqref{daX}, \eqref{daF} and \eqref{DAPHI} read
\begin{eqnarray}
i_W F=\mathcal{O}(\alpha^{'})~,~~~i_W d\Phi=\mathcal{O}(\alpha^{'2})~,~~~i_WX=\mathcal{O}(\alpha^{'2})
\end{eqnarray}
where we have denoted by $W$ the vector field dual to the 1-form $W=gdA$. Following an argument similar to that one in Section 3, we find
\begin{eqnarray}
\mathcal{L}_ W\Phi=\mathcal{O}(\alpha^{'2})~,~~~\mathcal{L}_W F=\mathcal{O}(\alpha^{'})~,~~~\mathcal{L}_W X=\mathcal{O}(\alpha^{'2})
\end{eqnarray}
Notice that \eqref{Hds2} and \eqref{M3metric} imply that
\begin{eqnarray}
H=\sqrt{\frac{k}{q}}\textrm{dvol}(AdS_3)+X+\mathcal{O}(\alpha^{'2})~.
\end{eqnarray}
Hence, we have found that warped $dS_2$ backgrounds in heterotic perturbation theory up to two loops are direct product manifolds of $AdS_3$ and an internal Riemannian manifold $M_7$. Such backgrounds have been classified in \cite{Beck:2015gqa}.

\section{$D=11$ $dS_3$ Killing superalgebra}

In the previous sections, we have shown that for heterotic warped product de Sitter geometries, the
warped product $dS_2$ solutions are $AdS_3\times M_7$ and all warped product $dS_n$ solutions for $n\ge 3$ are $\mathbb{R}^{1,n}\times  M_{9-n}$. Hence, de Sitter geometries are very restricted locally in heterotic supergravity. As is shown in \cite{DiGioia:2022bqg}, the conditions imposed by supersymmetry on $N=8$, $D=11$ warped $dS_4$ backgrounds are much weaker, and there is no foliation of $dS_4$ into $AdS_5$ or $\mathbb{R}^{1,4}$ appearing. In this section, we analyze warped $dS_3$ backgrounds in $D=11$ supergravity from the point of view of their Killing superalgebra $\mathfrak{g}$. In particular, we find that there is a rich algebraic structure arising from the super-Jacobi identities of $\mathfrak{g}$. This suggests that such backgrounds are not generically foliations of $dS_3$ into $AdS_4$ or $\mathbb{R}^{1,3}$. In this section we will refer to some formulae of section 2; from now on, it is understood that these formulae hold with $\alpha^{'}=0$.

Let us consider $dS_3\times_w M_8$ backgrounds in $D=11$ supergravity, where the metric is as in \eqref{EQ1}. Moreover, we assume the 4-form $F$ to be invariant under the isometry group $O(3,1)$ of $dS_3$, hence
\begin{eqnarray}
F=B\wedge\textrm{dvol}(dS_3)+X
\end{eqnarray}
where $B$ is a 1-form on $M_8$ and $X$ is a 4-form on $M_8$. Notice that $B$ and $X$ are closed forms on $M_8$, as a consequence of the Bianchi identities $dF=0$. We introduce on $M_{11}$ the co-frame \eqref{viel1}, where $a=3,4,\dots, \sharp$ and $y^{\alpha}$ are the co-ordinates on $M_8$. The Killing spinor equations of $D=11$ supergravity
\begin{eqnarray}
\bigg(\nabla_A-\frac{1}{288}\Gamma_A^{~~B_1\dots B_4}F_{B_1\dots B_4}+\frac{1}{36}F_{AB_1B_2B_3}\Gamma^{B_1B_2B_3}\bigg)\epsilon=0
\end{eqnarray}
can be integrated along $dS_3$, yielding \eqref{epsilonfinale}, where
\begin{eqnarray}
\mathcal{C}_{(3)}=-\frac{1}{2}\widetilde{\slashed{\nabla}}A+\frac{A}{288}\slashed{X}-\frac{1}{6}A^{-2}\slashed{B}\Gamma^{(3)}~.
\label{C3}
\end{eqnarray}
In \eqref{C3}, we have denoted $\Gamma^{(3)}:=\frac{1}{3!}\epsilon_{\mu_1\mu_2\mu_3}\Gamma^{\mu_1\mu_2\mu_3}$ the highest rank Gamma matrix on $dS_3$. Moreover, $\psi$ is a 32-component Majorana spinor on $M_8$ which satisfies \eqref{ksem9}, where
\begin{eqnarray}
\sigma_{a(3)}=\frac{1}{288}\Gamma_a^{~~b_1\dots b_4}X_{b_1\dots b_4}-\frac{1}{36}X_{ab_1b_2b_3}\Gamma^{b_1b_2b_3}+\frac{1}{12}A^{-3}\Gamma_a^{~~b}B_b\Gamma^{(3)}-\frac{1}{6}A^{-3}B_a\Gamma^{(3)}
\nonumber \\
\end{eqnarray}
supplemented by the quadratic conditions \eqref{ksemu26} and \eqref{ksem92}, where
\begin{eqnarray}
&&\widehat{\mathcal{C}}_{(3)}=\frac{1}{2}\widetilde{\slashed{\nabla}}A+\frac{A}{288}\slashed{X}+\frac{1}{6}A^{-2}\slashed{B}\Gamma^{(3)}~,
\nonumber \\
&&\widehat{\sigma}_{a(3)}=-\frac{1}{288}\Gamma_a^{~~b_1\dots b_4}X_{b_1\dots b_4}+\frac{1}{36}X_{ab_1b_2b_3}\Gamma^{b_1b_2b_3}+\frac{1}{12}A^{-3}\Gamma_a^{~~b}B_b\Gamma^{(3)}-\frac{1}{6}A^{-3}B_a\Gamma^{(3)}~.
\nonumber \\
\end{eqnarray}
Equations \eqref{ksem9} and \eqref{ksem92} imply that if $\psi$ satisfies \eqref{ksem9}, so do $\Gamma_{\mu}\mathcal{C}_{(3)}\psi$ and $\Gamma_{\mu\nu}\psi$. In the following it is useful to introduce the light-cone vielbein $\sqrt{2}\textbf{e}^{\pm}:=\textbf{e}^1\pm \textbf{e}^0$. In terms of this frame, the 11-dimensional metric reads
\begin{eqnarray}
ds^2(M_{11})=2\textbf{e}^+\textbf{e}^-+(\textbf{e}^2)^2+ds^2(M_8)~.
\end{eqnarray}
Without loss of generality, we take $\psi=\psi_+$, where $\Gamma_+\psi_+=0$. Let us assume there are $\mathfrak{n}$ linearly independent Killing spinors $\psi_+^r$, $r=1,2,\dots,\mathfrak{n}$, that is they satisfy \eqref{ksem9}. In other words, we take the vector space $V_+$ spanned by the Killing spinors $\psi_+^r$ to be $\mathfrak{n}$-dimensional. Then, since $\Gamma_2\mathcal{C}_{(3)}\psi_+^r$ are Killing spinors, it follows that
\begin{eqnarray}
\Gamma_2\mathcal{C}_{(3)}\psi_+^r=a^r_{~s}\psi_+^s
\label{acond}
\end{eqnarray}
where $a^r_{~s}$ is a constant $\mathfrak{n}\times \mathfrak{n}$ matrix. Using \eqref{acond} and the integrability conditions \eqref{ksemu26}, it is straightforward to show that
\begin{eqnarray}
a^2=-\frac{k}{4}\mathbb{I}_{\mathfrak{n}}
\label{a2}
\end{eqnarray}
hence, as $k>0$, it follows that $\textrm{dim}~V_{+}=\mathfrak{n}=2\mathfrak{m}$. Furthermore, equation \eqref{a2} implies that there exists a basis $\{\psi_+^{2\hat{r}-1},\psi_+^{2\hat{r}}~,~\hat{r}=1,2,\dots ,\mathfrak{m} \}$ of $V_{+}$ such that
\begin{eqnarray}
\Gamma_2\mathcal{C}_{(3)}\psi_+^{2\hat{r}-1}=\frac{\sqrt{k}}{2}\psi_+^{2\hat{r}}~,~~~~\Gamma_2\mathcal{C}_{(3)}\psi_+^{2\hat{r}}=-\frac{\sqrt{k}}{2}\psi_+^{2\hat{r}-1}~.
\label{simpli1}
\end{eqnarray}
The previous discussion implies that the linearly independent Killing spinors on $M_8$ are 
$\{\psi_{\pm}^{2\hat{r}}~,~\psi_{\pm}^{2\hat{r}-1}\}$, where $\psi_-^{2\hat{r}}$ and $\psi_-^{2\hat{r}-1}$ are given by
\begin{eqnarray}
\psi_-^{2\hat{r}-1}:=\Gamma_{-2}\psi_+^{2\hat{r}-1}~,~~~~~\psi_-^{2\hat{r}}:=\Gamma_{-2}\psi_+^{2\hat{r}}~.
\label{Bmeno}
\end{eqnarray}
Hence the number of supersymmetries $N$ preserved by warped $dS_3$ backgrounds in $D=11$ supergravity is $N=4\mathfrak{m}$ with $\mathfrak{m}=1,2,\dots,8$. In order to simplify the notation we denote $r:=\hat{r}$, hence in what follows $r=1,2,\dots ,\frac{\mathfrak{n}}{2}$. Using \eqref{epsilonfinale}, define
\begin{eqnarray}
&&\epsilon_1^{2r}:=\mathcal{U}^{-1/2}\big(1+x^{\mu}\Gamma_{\mu}\mathcal{C}_{(3)}\big)\psi_+^{2r}~,~~~~\epsilon_1^{2r-1}:=\mathcal{U}^{-1/2}\big(1+x^{\mu}\Gamma_{\mu}\mathcal{C}_{(3)}\big)\psi_+^{2r-1}
\nonumber \\
&&\epsilon_2^{2r}:=\mathcal{U}^{-1/2}\big(1+x^{\mu}\Gamma_{\mu}\mathcal{C}_{(3)}\big)\psi_-^{2r}~,~~~~\epsilon_2^{2r-1}:=\mathcal{U}^{-1/2}\big(1+x^{\mu}\Gamma_{\mu}\mathcal{C}_{(3)}\big)\psi_-^{2r-1}~.
\nonumber \\
\label{e1e2}
\end{eqnarray}
Inserting \eqref{simpli1} in \eqref{e1e2}, one obtains
\begin{eqnarray}
&&\epsilon_1^{2r}=\mathcal{U}^{-1/2}\big(\psi_+^{2r}-\frac{\sqrt{k}}{2}x^-\psi_-^{2r-1}-\frac{\sqrt{k}}{2}x^2\psi_+^{2r-1}\big)
\nonumber \\
&&\epsilon_1^{2r-1}=\mathcal{U}^{-1/2}\big(\psi_+^{2r-1}+\frac{\sqrt{k}}{2}x^-\psi_-^{2r}+\frac{\sqrt{k}}{2}x^2\psi_+^{2r}\big)
\nonumber \\
&&\epsilon_2^{2r}=\mathcal{U}^{-1/2}\big(\psi_-^{2r}-\sqrt{k}x^+\psi_+^{2r-1}+\frac{\sqrt{k}}{2}x^2\psi_-^{2r-1}\big)
\nonumber \\
&&\epsilon_2^{2r-1}=\mathcal{U}^{-1/2}\big(\psi_-^{2r-1}+\sqrt{k}x^+\psi_+^{2r}-\frac{\sqrt{k}}{2}x^2\psi_-^{2r}\big)~.
\label{spinorsds3}
\end{eqnarray}
The first step in constructing the $dS_3$ Killing superalgebra consists of computing the 11-dimensional bilinears 
\begin{eqnarray}
K(\epsilon_1,\epsilon_2):=\langle(\Gamma_+-\Gamma_-)\epsilon_1,\Gamma_A\epsilon_2\rangle \textbf{e}^A
\label{bilids3}
\end{eqnarray}
where $\epsilon_1$ and $\epsilon_2$ are Killing spinors. It is a general property of $D=11$ Killing superalgebras that the vector fields dual to \eqref{bilids3} - which will be denoted by the same name $K(\epsilon_1,\epsilon_2)$ - are Killing vectors \cite{Figueroa-OFarrill:2004qhu}, satisfying
\begin{eqnarray}
\mathcal{L}_{K(\epsilon_1,\epsilon_2)}g=0~,~~~~\mathcal{L}_{K(\epsilon_1,\epsilon_2)}F=0~.
\label{supgen}
\end{eqnarray}
In the following, we denote $K_1^{rs}:=K(\epsilon_1^r,\epsilon_1^s)$, $K_2^{rs}:=K(\epsilon_2^r,\epsilon_2^s)$ and $K_3^{rs}:=K(\epsilon_1^r,\epsilon_2^s)$. Notice that $K_1^{[rs]}=0$ and $K_2^{[rs]}=0$ as a consequence of \eqref{bilids3}. In the context of $AdS_3$ Killing superalgebras in $D=11$ supergravity, it can be shown via global arguments that the even sub-algebra $\mathfrak{g}_0$ of the Killing superalgebra $\mathfrak{g}=\mathfrak{g}_0\oplus \mathfrak{g}_1$ is  a direct sum of the isometry algebra of $AdS_3$ and the isometry algebra of the internal space \cite{Beck:2017wpm}. In the following, we argue that this is not the case for $dS_3$, that is a vector field $K$ satisfying \eqref{supgen} cannot be written as $K=A K^{\textrm{dS}}+K^{\textrm{int}}$, where $K^{\textrm{dS}}$ is a $dS_3$ Killing vector and $K^{\textrm{int}}$ is a Killing vector along $M_8$. Substituting \eqref{spinorsds3} in \eqref{bilids3}, a long computation shows that the bilinears are given by
\begin{eqnarray}
&&2K_1^{2r~2s}=V_1^{rs}+W_1^{rs}~,~~~~~~~~~~~~~~2K_2^{2r~2s}=V_3^{rs}+W_2^{rs}
\nonumber \\
&&2K_1^{2r-1~2s-1}=-V_1^{rs}+W_1^{rs}~,~~~~~~2K_2^{2r-1~2s-1}=-V_3^{rs}+W_2^{rs}
\nonumber \\
&&2K_1^{2r~2s-1}=V_2^{rs}-W_5^{rs}~,~~~~~~~~~~~2K_2^{2r~2s-1}=V_4^{rs}-W_6^{rs}
\label{bili1}
\end{eqnarray}
\begin{eqnarray}
&&2K_3^{(2r~2s)}=V_5^{rs}+W_3^{rs}~,~~~~~~~~~~~~~~2K_3^{[2r~2s]}=U_1^{rs}+W_8^{rs}
\nonumber \\
&&2K_3^{(2r-1~2s-1)}=-V_5^{rs}+W_3^{rs}~,~~~~~~2K_3^{[2r-1~2s-1]}=-U_1^{rs}+W_8^{rs}
\nonumber \\
&&2K_3^{(2r~2s-1)}=V_6^{rs}-W_7^{rs}~,~~~~~~~~~~~2K_3^{[2r~2s-1]}=U_2^{rs}+W_4^{rs}~.
\nonumber \\
\label{bili2}
\end{eqnarray}
In \eqref{bili1} and \eqref{bili2}, the vector fields $\{V_i^{rs}\}_{i=1}^6$ are defined by
\begin{eqnarray}
&&V_1^{rs}:=\sqrt{2k}~d^{rs}(L^{02}-L^{12})+\sqrt{2}~c^{rs}(W^0-W^1)
\nonumber \\
&&V_2^{rs}:=\sqrt{2}~d^{rs}(W^0-W^1)-\sqrt{2k}~c^{rs}(L^{02}-L^{12})
\nonumber \\
&&V_3^{rs}:=2\sqrt{2}~c^{rs}(W^0+W^1)-2\sqrt{2k}~d^{rs}(L^{02}+L^{12})
\nonumber \\
&&V_4^{rs}:=2\sqrt{2}~d^{rs}(W^0+W^1)+2\sqrt{2k}~c^{rs}(L^{02}+L^{12})
\nonumber \\
&&V_5^{rs}:=2\sqrt{k}~d^{rs}L^{01}+2c^{rs}W^2
\nonumber \\
&&V_6^{rs}:=-2\sqrt{k}~c^{rs}L^{01}+2d^{rs}W^2
\label{V12ds}
\end{eqnarray}
where  $c^{rs}$ and $d^{rs}$ are constant\footnote{The constancy of \eqref{crsdrs}, that is $\widetilde{\nabla}_a c^{rs}=\widetilde{\nabla}_a d^{rs}=0$, follows straightforwardly from the KSE \eqref{ksem9}.} symmetric $\frac{\mathfrak{n}}{2}\times\frac{\mathfrak{n}}{2}$ matrices given by
\begin{eqnarray}
&&c^{rs}:=A^{-1}\big(\langle\psi_+^{2r},\psi_+^{2s}\rangle-\langle\psi_+^{2r-1},\psi_+^{2s-1}\rangle\big)~,~~d^{rs}:=A^{-1}\big(\langle\psi_+^{2s},\psi_+^{2r-1}\rangle+\langle\psi_+^{2r},\psi_+^{2s-1}\rangle\big)
\nonumber \\
\label{crsdrs}
\end{eqnarray} 
and the vector fields $L^{\alpha\beta}$ and $W^{\alpha}$, with $\alpha,\beta=0,1,2$ are $dS_3$ Killing vector fields, defined by
\begin{eqnarray}
&&(L^{\alpha\beta})^{\mu}:=x^{\alpha}\eta^{\mu\beta}-x^{\beta}\eta^{\mu\alpha} ~,~~~(W^{\alpha})^{\mu}:=\big(1-\frac{k}{4}|x|^2\big)\eta^{\mu\alpha}+\frac{k}{2}x^{\alpha}x^{\mu}~.
\label{Wapp}
\end{eqnarray}
The vector fields \eqref{V12ds} are $dS_3$ Killing vectors, i.e.
\begin{eqnarray}
\mathcal{L}_{V_i^{rs}} g_{(3)}=0~~~~~~i=1,2,\dots, 6
\end{eqnarray}
where we have decomposed the 11-dimensional metric as $g=A^2 g_{(3)}+g_{(8)}$, where $g_{(3)}$ is the $dS_3$ metric and $g_{(8)}$ is the internal metric. Moreover, in \eqref{bili2}, the vector fields $\{U_i^{rs}\}_{i=1}^2$ are defined by
\begin{eqnarray}
&&U_1^{rs}:=2\bigg(\langle\psi_+^{2r},\Gamma^a\Gamma_2\psi_+^{2s}\rangle-\langle\psi_+^{2r-1},\Gamma^a\Gamma_2\psi_+^{2s-1}\rangle\bigg)\textbf{e}_a
\nonumber \\
&&U_2^{rs}:=2\bigg(\langle\psi_+^{2r},\Gamma^a\Gamma_2\psi_+^{2s-1}\rangle-\langle\psi_+^{2s},\Gamma^a\Gamma_2\psi_+^{2r-1}\rangle\bigg)\textbf{e}_a~.
\nonumber \\
\label{U2}
\end{eqnarray}
The vector fields \eqref{U2} are internal Killing vectors, i.e.
\begin{eqnarray}
\widetilde{\mathcal{L}}_{U_i^{rs}} g_{(8)}=0~~~~~i=1,2
\end{eqnarray}
which preserve the warp factor and the bosonic fluxes
\begin{eqnarray}
\widetilde{\mathcal{L}}_{U_i^{rs}} A=0~,~~~\widetilde{\mathcal{L}}_{U_i^{rs}} B=0~,~~~\widetilde{\mathcal{L}}_{U_i^{rs}} X=0~~~~~~~~i=1,2~.
\label{kint2}
\end{eqnarray}
The obstruction for $\mathfrak{g}_0$ to be a direct sum of the isometry algebra of $dS_3$ and the internal algebra is encoded in the vector fields $W_i^{rs}$ in \eqref{bili1} and \eqref{bili2}, which are defined by
\begin{eqnarray}
&&W_i^{rs}:=f^{rs}\xi_i+h_{i}~Z^{rs}~,~~~W_{i+4}^{rs}:=g^{rs}\xi_i+h_{i}~Y^{rs}~~~~~i=1,2,3,4~.
\label{allWS}
\end{eqnarray}
In \eqref{allWS}, $f^{rs}$ and $g^{rs}$ are functions on $M_8$ given by
\begin{eqnarray}
f^{rs}:=A^{-1}\big(\langle\psi_+^{2r},\psi_+^{2s}\rangle+\langle\psi_+^{2r-1},\psi_+^{2s-1}\rangle\big)~,~~~g^{rs}:=A^{-1}\big(\langle\psi_+^{2r-1},\psi_+^{2s}\rangle-\langle\psi_+^{2s-1},\psi_+^{2r}\rangle\big)~,
\nonumber \\
\label{grs}
\end{eqnarray}
$Z^{rs}$ and $Y^{rs}$ are vector fields on $M_8$ defined as
\begin{eqnarray}
&&Z^{rs}:=\bigg(\langle\psi_+^{2r},\Gamma^a\Gamma_2\psi_+^{2s-1}\rangle+\langle\psi_+^{2s},\Gamma^a\Gamma_2\psi_+^{2r-1}\rangle\bigg)\textbf{e}_a
\nonumber \\
&&Y^{rs}:=\bigg(\langle\psi_+^{2r},\Gamma^a\Gamma_2\psi_+^{2s}\rangle+\langle\psi_+^{2r-1},\Gamma^a\Gamma_2\psi_+^{2s-1}\rangle\bigg)\textbf{e}_a~.
\label{Yrs}
\end{eqnarray}
Moreover
\begin{eqnarray}
&&\xi_1:=-2\sqrt{2}(P^1-P^0)-\sqrt{2}(W^0-W^1)~,~~~\xi_2:=4\sqrt{2}(P^1+P^0)-2\sqrt{2}(W^0+W^1)
\nonumber \\
&&\xi_3:=4P^2-2W^2~,~~~~~~~~~~~~~~~~~~~~~~~~~~~~~~~~\xi_4 :=-2\sqrt{k}D
\label{xiss}
\end{eqnarray}
where in \eqref{xiss} $P^0$, $P^1$, $P^2$ and $D$ are $dS_3$ conformal Killing vectors given by
\begin{eqnarray}
&&(P^{\alpha})^{\mu}:=\eta^{\alpha\mu}~,~~~D^{\mu}:=x^{\mu}
\label{Dapp}
\end{eqnarray}
and\footnote{We define the light-cone co-ordinates $x^{\pm}:=\frac{1}{\sqrt{2}}(x^1\pm x^0)$. Notice that $|x|^2=2x^+x^-+(x^2)^2$.}
\begin{eqnarray}
&&h_1:=-2\mathcal{U}^{-1}\sqrt{k}x^-~,~~h_2:=4\mathcal{U}^{-1}\sqrt{k}x^+~,~~h_3:=2\mathcal{U}^{-1}\sqrt{k}x^2~,~~h_4:=2\mathcal{U}^{-1}\big(1-\frac{k}{4}|x|^2\big)~.
\nonumber \\
\label{KWHW}
\end{eqnarray}
Decomposing \eqref{supgen} along $g=A^2 g_{(3)}+g_{(8)}$ and using the integrability conditions \eqref{ksemu26}, we find the following conditions ($i=1,2,3,4$)
\begin{eqnarray}
&&\widetilde{\mathcal{L}}_{Z^{rs}}g_{(8)}=0~,~~~\widetilde{\mathcal{L}}_{Y^{rs}}g_{(8)}=0
\label{Killinginternal2}
\end{eqnarray}
\begin{eqnarray}
\mathcal{L}_{\xi_i}g_{(3)}=-2\sqrt{k}~h_i g_{(3)}~,~~~dh_i=\sqrt{k}\xi_i~,
\label{dhdh}
\end{eqnarray}
\begin{eqnarray}
&&A^{-1}\widetilde{\mathcal{L}}_{Z^{rs}}A=\sqrt{k}f^{rs}~,~~~A^{-1}\widetilde{\mathcal{L}}_{Y^{rs}}A=\sqrt{k}g^{rs}
\label{AA2}
\end{eqnarray}
\begin{eqnarray}
&&d f^{rs}=-\sqrt{k}A^{-2}Z^{rs}~,~~~d g^{rs}=-\sqrt{k}A^{-2}Y^{rs}
\label{dfdg2}
\end{eqnarray}
\begin{eqnarray}
&&\widetilde{\mathcal{L}}_{Z^{rs}}B=3\sqrt{k}f^{rs}B~,~~~\widetilde{\mathcal{L}}_{Y^{rs}}B=3\sqrt{k}g^{rs}B
\label{LIEB}
\end{eqnarray}
\begin{eqnarray}
&&\widetilde{\mathcal{L}}_{Z^{rs}}X=0~,~~~\widetilde{\mathcal{L}}_{Y^{rs}}X=0
\label{LIEX}
\end{eqnarray}
where in \eqref{dhdh}, we have denoted by $\xi_i$ the 1-form dual to the vector field $\xi_i$ \eqref{xiss}. In particular, equations \eqref{Killinginternal2}, \eqref{AA2}, \eqref{LIEB} and \eqref{LIEX} imply that $Z^{rs}$ and $Y^{rs}$ are internal Killing vector fields, which preserve $X$, but do not preserve either the warp factor $A$ or $B$. The appearance of the vector fields \eqref{allWS} in \eqref{bili1} and \eqref{bili2} implies that the even sub-algebra $\mathfrak{g}_0$ of the Killing superalgebra $\mathfrak{g}=\mathfrak{g}_0\oplus \mathfrak{g}_1$ is \textit{not} a direct sum of the isometry algebra of $dS_3$ and the isometry algebra of the internal space. Indeed, the vector fields \eqref{allWS} have components along the internal space $M_8$, which depend on the co-ordinates of $dS_3$. 

The next step in the construction of the $dS_3$ Killing superalgebra consists of computing the spinorial Lie derivatives 
\begin{eqnarray}
\mathcal{L}_K\epsilon=\nabla_K\epsilon+\frac{1}{8}\slashed{dK}\epsilon
\label{ddd}
\end{eqnarray}
where in \eqref{ddd} $K:=K(\epsilon_1,\epsilon_2)$ is the vector field dual to \eqref{bilids3}, $\nabla$ is the Levi-Civita connection on $M_{11}$ and $\epsilon$ is a Killing spinor. It is a general result of $D=11$ Killing superalgebras that the spinor \eqref{ddd} is Killing \cite{Figueroa-OFarrill:2004qhu}. In Appendix B we have listed all the spinorial Lie derivatives \eqref{ddd} and some additional algebraic conditions arising from \eqref{ddd}, see equations \eqref{liefirst}-\eqref{liefin}. Moreover, it is well known that the Killing superalgebra $\mathfrak{g}$ of $D=11$ supergravity satisfies the super-Jacobi identities \cite{Figueroa-OFarrill:2004qhu}
\begin{eqnarray}
[[x,y]_{\mathfrak{g}},z]_{\mathfrak{g}}+(-)^{|z|(|x|+|y|)}[[z,x]_{\mathfrak{g}},y]_{\mathfrak{g}}+(-)^{|x|(|y|+|z|)}[[y,z]_{\mathfrak{g}},x]_{\mathfrak{g}}=0
\label{sup2}
\end{eqnarray}
where $x,y,z\in\mathfrak{g}$, $|x|=0$ if $x\in\mathfrak{g}_0$, $|x|=1$ if $x\in\mathfrak{g}_1$ and
\begin{eqnarray}
[x,y]_{\mathfrak{g}}:=x\cdot y-(-)^{|x||y|}y\cdot x~.
\label{sup110}
\end{eqnarray}
Thus, it is important to check whether the super-Jacobi identities \eqref{sup2} of the $dS_3$ Killing superalgebra are automatically satisfied or provide some additional conditions. It turns out that there are indeed some non-trivial conditions arising from \eqref{sup2}. They are presented in Appendix B, see equations \eqref{ZU1one}-\eqref{g1g1g1sei}. A quite rich and complicated algebraic structure emerges, which strongly suggests that warped $dS_3$ backgrounds are not generically foliations into $AdS_4$ or $\mathbb{R}^{1,3}$. As a final comment, we remark that equations \eqref{u1u1uno}-\eqref{u2u2due} are reminiscent of the quadratic conditions obeyed by the structure constants of a (complex) 3-algebra \cite{Bagger:2008se}; however, by carefully comparing \eqref{u1u1uno}-\eqref{u2u2due} with equation (2) of  \cite{Bagger:2008se}, we find that the index structure does not match, hence there is no simple interpretation of $D=11$ $dS_3$ Killing superalgebras in terms of 3-algebras.

\section{Conclusion}

In this work, we have classified all warped product $dS_n$ solutions of heterotic perturbation theory, up to two loops. We have found that
\begin{itemize}

\item{} For $n\ge 3$, all such solutions are $\mathbb{R}^{1,n-1}\times M_{9-n}$, where $M_{9-n}$ is a $(9-n)$-dimensional Riemannian manifold. Moreover, the 1-form $d\Phi$, the 3-form $H$ and the non-abelian field strength $F$ have non-vanishing components only along $M_{9-n}$. Backgrounds of these form have been studied in \cite{Ivanov} and \cite{Ivanov2}.

\item{} For $n=2$, all such solutions are $AdS_3\times M_7$, where $M_7$ is a 7-dimensional Riemannian manifold. These backgrounds have been classified in \cite{Beck:2015gqa}.

\end{itemize}
The analysis done in this paper can be straightforwardly adapted to investigate the case $k=0$, which corresponds to warped product Minkowski solutions $\mathbb{R}^{1,n-1}\times_w M_{10-n}$. Notice that such backgrounds can be thought of as the limit $\alpha^{'}\to 0$ of \eqref{EQ1} and \eqref{dSD} when $k^0=0$, that is $k=k^1\alpha^{'}+\mathcal{O}(\alpha^{'2})$. We find that warped product $\mathbb{R}^{1,n-1}$ Minkowski backgrounds are $\mathbb{R}^{1,n-1}\times M_{10-n}$, for $n \ge 3$. Moreover, in the case $n=2$, there are two different classes: 
\begin{itemize}

\item{} $AdS_3\times M_7$, where $M_7$ is a 7-dimensional Riemannian manifold.

\item{} near-horizon heterotic black hole solutions, with $dh=0$, where $h=-d\log A^2$ \cite{Gutowski:2009wm}, \cite{Gran:2011ak}.

\end{itemize}
Perturbative $\alpha{'}$ corrections to the cosmological constant $k$ have also been studied in the context of warped heterotic compactifications. In \cite{Green:2011cn}, assuming the internal manifold to be compact and without boundary, it is shown that de-Sitter backgrounds are excluded. The result is established taking into account leading $\alpha^{'}$ effects at tree-level and does not utilize supersymmetry. Moreover, in \cite{Gautason:2012tb} it is found that heterotic string compactifications at tree-level yields 4-dimensional Minkowski as the only maximally supersymmetric solution, to all orders in $\alpha^{'}$. The work in \cite{Green:2011cn, Gautason:2012tb} is extended and complemented in \cite{Kutasov:2015eba}, where, using a world-sheet CFT argument, it is shown that the heterotic string essentially rules out $dS_n$ vacua with $n\ge 4$, to all orders in $\alpha^{'}$.

In Section 6, we have shown that warped product $dS_3$ backgrounds in $D=11$ supergravity are less restricted than in heterotic supergravity. Indeed, by analyzing the Killing superalgebra of these backgrounds, a rich algebraic structure emerges, which suggests that such backgrounds are not generically foliations into $AdS_4$ or $\mathbb{R}^{1,4}$. A similar situation occurs for warped $dS_4$ backgrounds in $D=11$ supergravity: indeed, in \cite{DiGioia:2022bqg}, by analyzing the constraints on the geometry imposed by minimal $N=8$ supersymmetry, it is shown that warped $dS_4$ backgrounds are not generically foliations into $AdS_5$ or $\mathbb{R}^{1,5}$.

\setcounter{section}{0}
\setcounter{subsection}{0}

\appendix{Curvature: conventions and useful formulae}

In this Appendix, we list our conventions for the curvature tensor of a connection. The curvature of a generic connection $\Gamma$ is given by \cite{Fontanella:2016aok}
\begin{eqnarray}
R_{AB,~~D}^{~~~~C}=\textbf{e}_A \Gamma^C_{~BD}-\textbf{e}_B\Gamma^C_{~AD}+\Gamma^C_{~AN}\Gamma^N_{~BD}-\Gamma^C_{~BN}\Gamma^N_{~AD}~.
\label{Rconngen}
\end{eqnarray}
Let $\Gamma$ be the Levi-Civita connection of the 10-dimensional spacetime  \eqref{11DM}. Then, the non-vanishing components of the spin connection in the frame \eqref{viel1} are given by
\begin{eqnarray}
\Omega_{\mu,\nu\rho}&=&kA^{-1}x_{[\nu}\eta_{\rho]\mu}
\nonumber \\
\Omega_{\mu,a\nu}&=&-\eta_{\mu\nu}A^{-1}{\tilde{\nabla}}_a A
\nonumber \\
\Omega_{a,bc}&=&\widetilde{\Omega}_{a,bc}
\label{spinconnection2}
\end{eqnarray}
where $\widetilde{\Omega}_{a,bc}$ is the spin connection on $M_{10-n}$. Utilizing  \eqref{Rconngen} and \eqref{spinconnection2}, the non-vanishing components of the 10-dimensional Riemann tensor are 
\begin{eqnarray}
R_{\mu\nu\rho\sigma}&=&\big(\eta_{\mu\rho}\eta_{\nu\sigma}-\eta_{\mu\sigma}\eta_{\nu\rho}\big)A^{-2}\big(k-(\widetilde{\nabla}A)^2\big)
\nonumber \\
R_{a\mu b\nu}&=&-\eta_{\mu\nu}A^{-1}\widetilde{\nabla}_a\widetilde{\nabla}_bA
\nonumber \\
R_{abcd}&=&\widetilde{R}_{abcd}
\label{Riemann}
\end{eqnarray}
where $\widetilde{R}_{abcd}$ denotes the Riemann tensor on $M_{10-n}$.  Implementing \eqref{Riemann}, we compute the 10-dimensional Ricci tensor, whose non-vanishing components are 
\begin{eqnarray}
R_{\mu\nu}&=&\eta_{\mu\nu}\bigg(k(n-1)A^{-2}-A^{-1}\widetilde{\nabla}^a\widetilde{\nabla}_aA-(n-1)A^{-2}(\widetilde{\nabla}A)^2\bigg)
\nonumber \\
R_{ab}&=&\widetilde{R}_{ab}-nA^{-1}\widetilde{\nabla}_a\widetilde{\nabla}_b A
\label{Ricci}
\end{eqnarray}
where $\widetilde{R}_{ab}$ is the Ricci tensor on $M_{10-n}$. \\
\indent
Let us introduce the connections with torsion $\hat{\nabla}$ and $\check{\nabla}$ 
\begin{eqnarray}
\check{\nabla}_M\xi^N=\nabla_M\xi^N-\frac{1}{2}H^N_{~~ML}\xi^L~,~~~~\hat{\nabla}_M\xi^N=\nabla_M\xi^N+\frac{1}{2}H^N_{~~ML}\xi^L
\label{hatnabla}
\end{eqnarray}
where $\nabla$ is the Levi-Civita connection and $\xi$ is a vector field. Using \eqref{Rconngen} and \eqref{hatnabla}, it follows that
\begin{eqnarray}
\check{R}_{AB,CD}=R_{ABCD}-\frac{1}{2}\nabla_AH_{CBD}+\frac{1}{2}\nabla_B H_{CAD}+\frac{1}{4}H_{CAN}H^N_{~~BD}-\frac{1}{4}H_{CBN}H^N_{~~AD}
\nonumber \\
\label{checkR}
\end{eqnarray}
and
\begin{eqnarray}
\hat{R}_{AB,CD}=R_{ABCD}+\frac{1}{2}\nabla_AH_{CBD}-\frac{1}{2}\nabla_B H_{CAD}+\frac{1}{4}H_{CAN}H^N_{~~BD}-\frac{1}{4}H_{CBN}H^N_{~~AD}~.
\nonumber \\
\label{hatR}
\end{eqnarray}
Notice that $\check{R}_{AB,CD}$ and $\hat{R}_{AB,CD}$ are anti-symmetric in the first two indices and the last two indices. Moreover, equations \eqref{checkR} and \eqref{hatR} imply the identity
\begin{eqnarray}
\check{R}_{AB,CD}-\hat{R}_{CD,AB}=\frac{1}{2}(dH)_{ABCD}~.
\label{identity}
\end{eqnarray}
Furthermore, notice that the integrability conditions of the gravitino KSE \eqref{KSEcorr1}
\begin{eqnarray}
[\hat{\nabla}_M,\hat{\nabla}_N]\epsilon=\mathcal{O}(\alpha^{'2})
\end{eqnarray}
can be expressed in terms of $\hat{R}$ as follows
\begin{eqnarray}
\hat{R}_{MN,RS}\Gamma^{RS}\epsilon=\mathcal{O}(\alpha^{'2})~.
\label{intuse}
\end{eqnarray}
Using \eqref{identity}, we can rewrite \eqref{intuse} as
\begin{eqnarray}
\bigg(\check{R}_{RS,MN}-\frac{1}{2}(dH)_{MNRS}\bigg)\Gamma^{RS}\epsilon=\mathcal{O}(\alpha^{'2})~.
\label{intuse2}
\end{eqnarray}
Since $dH=\mathcal{O}(\alpha^{'})$ by means of the Bianchi identities \eqref{BianchiDH}, it follows that \eqref{intuse2} is equivalent to
\begin{eqnarray}
\check{R}_{RS,MN}\Gamma^{RS}\epsilon=\mathcal{O}(\alpha^{'})
\end{eqnarray}
which is the same condition as that satisfied by the curvature of the gauge sector $F$ in the gaugino KSE \eqref{KSEcorr3}.

\appendix{$dS_3$ superalgebras}

Let us start this appendix by listing the spinorial Lie derivatives \eqref{ddd}. Using \eqref{spinorsds3}, \eqref{V12ds}, \eqref{U2} and \eqref{allWS}, a lengthy computation yields

\begin{eqnarray}
&&\mathcal{L}_{V_1^{mn}}\epsilon_1^{2r}=0~,~~~~~~~~~~~~~~~~~~~~~~~~~~~~~~~~~~~~~~\mathcal{L}_{V_1^{mn}}\epsilon_1^{2r-1}=0
\nonumber \\
&&\mathcal{L}_{V_1^{mn}}\epsilon_2^{2r}=2\sqrt{k}~d^{mn}\epsilon_1^{2r}+2\sqrt{k}~c^{mn}\epsilon_1^{2r-1}~,~~~\mathcal{L}_{V_1^{mn}}\epsilon_2^{2r-1}=2\sqrt{k}~d^{mn}\epsilon_1^{2r-1}-2\sqrt{k}~c^{mn}\epsilon_1^{2r}
\nonumber \\
\label{liefirst}
\end{eqnarray}

\begin{eqnarray}
&&\mathcal{L}_{V_2^{mn}}\epsilon_1^{2r}=0~,~~~~~~~~~~~~~~~~~~~~~~~~~~~~~~~~~~~~~~\mathcal{L}_{V_2^{mn}}\epsilon_1^{2r-1}=0
\nonumber \\
&&\mathcal{L}_{V_2^{mn}}\epsilon_2^{2r}=2\sqrt{k}~d^{mn}\epsilon_1^{2r-1}-2\sqrt{k}~c^{mn}\epsilon_1^{2r}~,~~~\mathcal{L}_{V_2^{mn}}\epsilon_2^{2r-1}=-2\sqrt{k}~d^{mn}\epsilon_1^{2r}-2\sqrt{k}~c^{mn}\epsilon_1^{2r-1}
\nonumber \\
\end{eqnarray}

\begin{eqnarray}
&&\mathcal{L}_{V_3^{mn}}\epsilon_1^{2r}=-2\sqrt{k}~c^{mn}\epsilon_2^{2r-1}-2\sqrt{k}~d^{mn}\epsilon_2^{2r}~,~~~\mathcal{L}_{V_3^{mn}}\epsilon_1^{2r-1}=2\sqrt{k}~c^{mn}\epsilon_2^{2r}-2\sqrt{k}~d^{mn}\epsilon_2^{2r-1}
\nonumber \\
&&\mathcal{L}_{V_3^{mn}}\epsilon_2^{2r}=0~,~~~~~~~~~~~~~~~~~~~~~~~~~~~~~~~~~~~~~~~~~\mathcal{L}_{V_3^{mn}}\epsilon_2^{2r-1}=0
\end{eqnarray}

\begin{eqnarray}
&&\mathcal{L}_{V_4^{mn}}\epsilon_1^{2r}=-2\sqrt{k}~d^{mn}\epsilon_2^{2r-1}+2\sqrt{k}~c^{mn}\epsilon_2^{2r}~,~~~\mathcal{L}_{V_4^{mn}}\epsilon_1^{2r-1}=2\sqrt{k}~d^{mn}\epsilon_2^{2r}+2\sqrt{k}~c^{mn}\epsilon_2^{2r-1}
\nonumber \\
&&\mathcal{L}_{V_4^{mn}}\epsilon_2^{2r}=0~,~~~~~~~~~~~~~~~~~~~~~~~~~~~~~~~~~~~~~~~~~\mathcal{L}_{V_4^{mn}}\epsilon_2^{2r-1}=0
\end{eqnarray}

\begin{eqnarray}
&&\mathcal{L}_{V_5^{mn}}\epsilon_1^{2r}=-\sqrt{k}~d^{mn}\epsilon_1^{2r}-\sqrt{k}~c^{mn}\epsilon_1^{2r-1}~,~\mathcal{L}_{V_5^{mn}}\epsilon_1^{2r-1}=-\sqrt{k}~d^{mn}\epsilon_1^{2r-1}+\sqrt{k}~c^{mn}\epsilon_1^{2r}
\nonumber \\
&&\mathcal{L}_{V_5^{mn}}\epsilon_2^{2r}=\sqrt{k}~d^{mn}\epsilon_2^{2r}+\sqrt{k}~c^{mn}\epsilon_2^{2r-1}~,~~~\mathcal{L}_{V_5^{mn}}\epsilon_2^{2r-1}=\sqrt{k}~d^{mn}\epsilon_2^{2r-1}-\sqrt{k}~c^{mn}\epsilon_2^{2r}
\nonumber \\
\end{eqnarray}

\begin{eqnarray}
&&\mathcal{L}_{V_6^{mn}}\epsilon_1^{2r}=\sqrt{k}~c^{mn}\epsilon_1^{2r}-\sqrt{k}~d^{mn}\epsilon_1^{2r-1}~,~~~\mathcal{L}_{V_6^{mn}}\epsilon_1^{2r-1}=\sqrt{k}~c^{mn}\epsilon_1^{2r-1}+\sqrt{k}~d^{mn}\epsilon_1^{2r}
\nonumber \\
&&\mathcal{L}_{V_6^{mn}}\epsilon_2^{2r}=-\sqrt{k}~c^{mn}\epsilon_2^{2r}+\sqrt{k}~d^{mn}\epsilon_2^{2r-1}~,~\mathcal{L}_{V_6^{mn}}\epsilon_2^{2r-1}=-\sqrt{k}~c^{mn}\epsilon_2^{2r-1}-\sqrt{k}~d^{mn}\epsilon_2^{2r}
\nonumber \\
\label{spinds36}
\end{eqnarray}
where $c^{mn}$ and $d^{mn}$ are the constants given by \eqref{crsdrs}. Moreover
\begin{eqnarray}
&&\mathcal{L}_{U_i^{mn}}\epsilon_1^{2r}=(\alpha_i)^{mnr}_{~~~~~p}\epsilon_1^{2p}+(\beta_i)^{mnr}_{~~~~~p}\epsilon_1^{2p-1}~,~~~\mathcal{L}_{U_i^{mn}}\epsilon_1^{2r-1}=-(\beta_i)^{mnr}_{~~~~~p}\epsilon_1^{2p}+(\alpha_i)^{mnr}_{~~~~~p}\epsilon_1^{2p-1}
\nonumber \\
&&\mathcal{L}_{U_i^{mn}}\epsilon_2^{2r}=(\alpha_i)^{mnr}_{~~~~~p}\epsilon_2^{2p}+(\beta_i)^{mnr}_{~~~~~p}\epsilon_2^{2p-1}~,~~~\mathcal{L}_{U_i^{mn}}\epsilon_2^{2r-1}=-(\beta_i)^{mnr}_{~~~~~p}\epsilon_2^{2p}+(\alpha_i)^{mnr}_{~~~~~p}\epsilon_2^{2p-1}
\nonumber \\
\label{spinorialkint4}
\end{eqnarray}
where $i=1,2$ and $(\alpha_i)^{mnr}_{~~~~~p}, (\beta_i)^{mnr}_{~~~~~p} $ are constants which are anti-symmetric in $m,n$. Equations \eqref{spinorialkint4} yield
\begin{eqnarray}
&&\widetilde{\mathcal{L}}_{U_i^{mn}}\psi_+^{2r}=(\alpha_i)^{mnr}_{~~~~~p}\psi_+^{2p}+(\beta_i)^{mnr}_{~~~~~p}\psi_+^{2p-1}~,~~~\widetilde{\mathcal{L}}_{U_i^{mn}}\psi_+^{2r-1}=(\alpha_i)^{mnr}_{~~~~~p}\psi_+^{2p-1}-(\beta_i)^{mnr}_{~~~~~p}\psi_+^{2p}~.
\nonumber \\
\label{spinspin}
\end{eqnarray}
Moreover
\begin{eqnarray}
&&\mathcal{L}_{W_1^{mn}}\epsilon_1^{2r}=\mathcal{L}_{W_1^{mn}}\epsilon_1^{2r-1}=\mathcal{L}_{W_2^{mn}}\epsilon_2^{2r}=\mathcal{L}_{W_2^{mn}}\epsilon_2^{2r-1}=0
\nonumber \\
&&\mathcal{L}_{W_1^{mn}}\epsilon_2^{2r}=A^{mnr}_{~~~~~p}\epsilon_1^{2p}+B^{mnr}_{~~~~~p}\epsilon_1^{2p-1}~,~~~~~\mathcal{L}_{W_1^{mn}}\epsilon_2^{2r-1}=B^{mnr}_{~~~~~p}\epsilon_1^{2p}-A^{mnr}_{~~~~~p}\epsilon_1^{2p-1}
\nonumber \\
&&\mathcal{L}_{W_2^{mn}}\epsilon_1^{2r}=-A^{mnr}_{~~~~~p}\epsilon_2^{2p}-B^{mnr}_{~~~~~p}\epsilon_2^{2p-1}~,~~~\mathcal{L}_{W_2^{mn}}\epsilon_1^{2r-1}=-B^{mnr}_{~~~~~p}\epsilon_2^{2p}+A^{mnr}_{~~~~~p}\epsilon_2^{2p-1}
\nonumber \\
\label{LIEWIN}
\end{eqnarray}
\begin{eqnarray}
&&2\mathcal{L}_{W_3^{mn}}\epsilon_1^{2r}=-A^{mnr}_{~~~~~p}\epsilon_1^{2p}-B^{mnr}_{~~~~~p}\epsilon_1^{2p-1}~,~~~2\mathcal{L}_{W_3^{mn}}\epsilon_1^{2r-1}=-B^{mnr}_{~~~~~p}\epsilon_1^{2p}+A^{mnr}_{~~~~~p}\epsilon_1^{2p-1}
\nonumber \\
&&2\mathcal{L}_{W_3^{mn}}\epsilon_2^{2r}=A^{mnr}_{~~~~~p}\epsilon_2^{2p}+B^{mnr}_{~~~~~p}\epsilon_2^{2p-1}~,~~~~~2\mathcal{L}_{W_3^{mn}}\epsilon_2^{2r-1}=B^{mnr}_{~~~~~p}\epsilon_2^{2p}-A^{mnr}_{~~~~~p}\epsilon_2^{2p-1}
\nonumber \\
&&2\mathcal{L}_{W_4^{mn}}\epsilon_1^{2r}=-B^{mnr}_{~~~~~p}\epsilon_1^{2p}+A^{mnr}_{~~~~~p}\epsilon_1^{2p-1}~,~~~2\mathcal{L}_{W_4^{mn}}\epsilon_1^{2r-1}=A^{mnr}_{~~~~~p}\epsilon_1^{2p}+B^{mnr}_{~~~~~p}\epsilon_1^{2p-1}
\nonumber \\
&&2\mathcal{L}_{W_4^{mn}}\epsilon_2^{2r}=-B^{mnr}_{~~~~~p}\epsilon_2^{2p}+A^{mnr}_{~~~~~p}\epsilon_2^{2p-1}~,~~~2\mathcal{L}_{W_4^{mn}}\epsilon_2^{2r-1}=A^{mnr}_{~~~~~p}\epsilon_2^{2p}+B^{mnr}_{~~~~~p}\epsilon_2^{2p-1}
\nonumber \\
\end{eqnarray}

\begin{eqnarray}
&&\mathcal{L}_{W_5^{mn}}\epsilon_1^{2r}=\mathcal{L}_{W_5^{mn}}\epsilon_1^{2r-1}=\mathcal{L}_{W_6^{mn}}\epsilon_2^{2r}=\mathcal{L}_{W_6^{mn}}\epsilon_2^{2r-1}=0
\nonumber \\
&&\mathcal{L}_{W_5^{mn}}\epsilon_2^{2r}=F^{mnr}_{~~~~~p}\epsilon_1^{2p}+G^{mnr}_{~~~~~p}\epsilon_1^{2p-1}~,~~~~~\mathcal{L}_{W_5^{mn}}\epsilon_2^{2r-1}=G^{mnr}_{~~~~~p}\epsilon_1^{2p}-F^{mnr}_{~~~~~p}\epsilon_1^{2p-1}
\nonumber \\
&&\mathcal{L}_{W_6^{mn}}\epsilon_1^{2r}=-F^{mnr}_{~~~~~p}\epsilon_2^{2p}-G^{mnr}_{~~~~~p}\epsilon_2^{2p-1}~,~~~\mathcal{L}_{W_6^{mn}}\epsilon_1^{2r-1}=-G^{mnr}_{~~~~~p}\epsilon_2^{2p}+F^{mnr}_{~~~~~p}\epsilon_2^{2p-1}
\nonumber \\
\end{eqnarray}

\begin{eqnarray}
&&2\mathcal{L}_{W_7^{mn}}\epsilon_1^{2r}=-F^{mnr}_{~~~~~p}\epsilon_1^{2p}-G^{mnr}_{~~~~~p}\epsilon_1^{2p-1}~,~~~2\mathcal{L}_{W_7^{mn}}\epsilon_1^{2r-1}=-G^{mnr}_{~~~~~p}\epsilon_1^{2p}+F^{mnr}_{~~~~~p}\epsilon_1^{2p-1}
\nonumber \\
&&2\mathcal{L}_{W_7^{mn}}\epsilon_2^{2r}=F^{mnr}_{~~~~~p}\epsilon_2^{2p}+G^{mnr}_{~~~~~p}\epsilon_2^{2p-1}~,~~~~~2\mathcal{L}_{W_7^{mn}}\epsilon_2^{2r-1}=G^{mnr}_{~~~~~p}\epsilon_2^{2p}-F^{mnr}_{~~~~~p}\epsilon_2^{2p-1}
\nonumber \\
&&2\mathcal{L}_{W_8^{mn}}\epsilon_1^{2r}=-G^{mnr}_{~~~~~p}\epsilon_1^{2p}+F^{mnr}_{~~~~~p}\epsilon_1^{2p-1}~,~~~2\mathcal{L}_{W_8^{mn}}\epsilon_1^{2r-1}=F^{mnr}_{~~~~~p}\epsilon_1^{2p}+G^{mnr}_{~~~~~p}\epsilon_1^{2p-1}
\nonumber \\
&&2\mathcal{L}_{W_8^{mn}}\epsilon_2^{2r}=-G^{mnr}_{~~~~~p}\epsilon_2^{2p}+F^{mnr}_{~~~~~p}\epsilon_2^{2p-1}~,~~~2\mathcal{L}_{W_8^{mn}}\epsilon_2^{2r-1}=F^{mnr}_{~~~~~p}\epsilon_2^{2p}+G^{mnr}_{~~~~~p}\epsilon_2^{2p-1}
\nonumber \\
\label{LIEWFIN}
\end{eqnarray}
where $A^{pqr}_{~~~~l}$, $B^{pqr}_{~~~~l}$, $F^{pqr}_{~~~~l}$ and $G^{pqr}_{~~~~l}$ are constants. In particular $A^{pqr}_{~~~~l}$ and $B^{pqr}_{~~~~l}$ are symmetric in $p,q$ while $F^{pqr}_{~~~~l}$ and $G^{pqr}_{~~~~l}$ are anti-symmetric in $p,q$. Moreover
\begin{eqnarray}
&&4\widetilde{\mathcal{L}}_{Z^{mn}}\psi_+^{2r}=A^{mnr}_{~~~~~p}\psi_+^{2p-1}-B^{mnr}_{~~~~~p}\psi_+^{2p}~,~~~4\widetilde{\mathcal{L}}_{Z^{mn}}\psi_+^{2r-1}=A^{mnr}_{~~~~~p}\psi_+^{2p}+B^{mnr}_{~~~~~p}\psi_+^{2p-1}
\nonumber \\
&&4\widetilde{\mathcal{L}}_{Y^{mn}}\psi_+^{2r}=F^{mnr}_{~~~~~p}\psi_+^{2p-1}-G^{mnr}_{~~~~~p}\psi_+^{2p}~,~~~4\widetilde{\mathcal{L}}_{Y^{mn}}\psi_+^{2r-1}=F^{mnr}_{~~~~~p}\psi_+^{2p}+G^{mnr}_{~~~~~p}\psi_+^{2p-1}
\nonumber \\
\label{spinspin4}
\end{eqnarray}
Furthermore, the following algebraic conditions hold
\begin{eqnarray}
&&2\sqrt{k}\big(f^{mn}\psi_+^{2r-1}+A^{-1}\slashed{Z}^{mn}\Gamma_2\psi_+^{2r}\big)=A^{mnr}_{~~~~~p}\psi_+^{2p}+B^{mnr}_{~~~~~p}\psi_+^{2p-1}
\nonumber \\
&&2\sqrt{k}\big(f^{mn}\psi_+^{2r}-A^{-1}\slashed{Z}^{mn}\Gamma_2\psi_+^{2r-1}\big)=A^{mnr}_{~~~~~p}\psi_+^{2p-1}-B^{mnr}_{~~~~~p}\psi_+^{2p}
\nonumber \\
&&2\sqrt{k}\big(g^{mn}\psi_+^{2r-1}+A^{-1}\slashed{Y}^{mn}\Gamma_2\psi_+^{2r}\big)=F^{mnr}_{~~~~~p}\psi_+^{2p}+G^{mnr}_{~~~~~p}\psi_+^{2p-1}
\nonumber \\
&&2\sqrt{k}\big(g^{mn}\psi_+^{2r}-A^{-1}\slashed{Y}^{mn}\Gamma_2\psi_+^{2r-1}\big)=F^{mnr}_{~~~~~p}\psi_+^{2p-1}-G^{mnr}_{~~~~~p}\psi_+^{2p}
\nonumber \\
\label{liefin}
\end{eqnarray}
In the final part of this Appendix, we list the Super-Jacobi identities \eqref{sup2} of the $dS_3$ Killing superalgebra. Using \eqref{spinorsds3}, \eqref{V12ds}, \eqref{U2} and \eqref{allWS}, we find the following results

\begin{itemize}

\item{$(\mathfrak{g}_0,\mathfrak{g}_0,\mathfrak{g}_0)$}: these conditions, which correspond to the ordinary Jacobi identities of a Lie algebra, are automatically satisfied.

\item{$(\mathfrak{g}_0,\mathfrak{g}_0,\mathfrak{g}_1)$} the non-trivial conditions are given by

\begin{eqnarray}
&&-A^{mn[p}_{~~~~~~l}B^{q]lr}_{~~~~t}+B^{mn[p}_{~~~~~~l}G^{q]lr}_{~~~~t}-(\alpha_1)^{pqr}_{~~~~l}B^{mnl}_{~~~~t}+(\beta_1)^{pqr}_{~~~~l}A^{mnl}_{~~~~t}
\nonumber \\
&&+A^{mnr}_{~~~~l}(\beta_1)^{pql}_{~~~~t}+B^{mnr}_{~~~~l}(\alpha_1)^{pql}_{~~~~t}=0~,
\label{ZU1one}
\end{eqnarray}
\begin{eqnarray}
&&A^{mn[p}_{~~~~~~l}A^{q]lr}_{~~~~t}-B^{mn[p}_{~~~~~~l}F^{q]lr}_{~~~~t}+(\alpha_1)^{pqr}_{~~~~l}A^{mnl}_{~~~~t}+(\beta_1)^{pqr}_{~~~~l}B^{mnl}_{~~~~t}
\nonumber \\
&&-A^{mnr}_{~~~~l}(\alpha_1)^{pql}_{~~~~t}+B^{mnr}_{~~~~l}(\beta_1)^{pql}_{~~~~t}=0~,
\label{ZU1two}
\end{eqnarray}

\begin{eqnarray}
&&-A^{mn[p}_{~~~~~~l}G^{q]lr}_{~~~~t}-B^{mn[p}_{~~~~~~l}B^{q]lr}_{~~~~t}-(\alpha_2)^{pqr}_{~~~~l}B^{mnl}_{~~~~t}+(\beta_2)^{pqr}_{~~~~l}A^{mnl}_{~~~~t}
\nonumber \\
&&+A^{mnr}_{~~~~l}(\beta_2)^{pql}_{~~~~t}+B^{mnr}_{~~~~l}(\alpha_2)^{pql}_{~~~~t}=0~,
\label{ZU2one}
\end{eqnarray}
\begin{eqnarray}
&&A^{mn[p}_{~~~~~~l}F^{q]lr}_{~~~~t}+B^{mn[p}_{~~~~~~l}A^{q]lr}_{~~~~t}+(\beta_2)^{pqr}_{~~~~l}B^{mnl}_{~~~~t}-A^{mnr}_{~~~~l}(\alpha_2)^{pql}_{~~~~t}
\nonumber \\
&&+(\alpha_2)^{pqr}_{~~~~l}A^{mnl}_{~~~~~~t}+B^{mnr}_{~~~~~~l}(\beta_2)^{pql}_{~~~~t}=0~,
\label{ZU2two}
\end{eqnarray}

\begin{eqnarray}
&&-F^{mn[p}_{~~~~~~l}B^{q]lr}_{~~~~t}+G^{mn[p}_{~~~~~~l}G^{q]lr}_{~~~~t}-(\alpha_1)^{pqr}_{~~~~l}G^{mnl}_{~~~~t}+(\beta_1)^{pqr}_{~~~~l}F^{mnl}_{~~~~t}
\nonumber \\
&&+F^{mnr}_{~~~~l}(\beta_1)^{pql}_{~~~~t}+G^{mnr}_{~~~~l}(\alpha_1)^{pql}_{~~~~t}=0~,
\label{YU1one}
\end{eqnarray}
\begin{eqnarray}
&&F^{mn[p}_{~~~~~~l}A^{q]lr}_{~~~~t}-G^{mn[p}_{~~~~~~l}F^{q]lr}_{~~~~t}+(\alpha_1)^{pqr}_{~~~~l}F^{mnl}_{~~~~t}+(\beta_1)^{pqr}_{~~~~l}G^{mnl}_{~~~~t}
\nonumber \\
&&-F^{mnr}_{~~~~l}(\alpha_1)^{pql}_{~~~~t}+G^{mnr}_{~~~~l}(\beta_1)^{pql}_{~~~~t}=0~,
\label{YU1two}
\end{eqnarray}

\begin{eqnarray}
&&-F^{mn[p}_{~~~~~~l}G^{q]lr}_{~~~~t}-G^{mn[p}_{~~~~~~l}B^{q]lr}_{~~~~t}-(\alpha_2)^{pqr}_{~~~~l}G^{mnl}_{~~~~t}+(\beta_2)^{pqr}_{~~~~l}F^{mnl}_{~~~~t}
\nonumber \\
&&+F^{mnr}_{~~~~l}(\beta_2)^{pql}_{~~~~t}+G^{mnr}_{~~~~l}(\alpha_2)^{pql}_{~~~~t}=0~,
\label{YU2one}
\end{eqnarray}
\begin{eqnarray}
&&F^{mn[p}_{~~~~~~l}F^{q]lr}_{~~~~t}+G^{mn[p}_{~~~~~~l}A^{q]lr}_{~~~~t}+(\alpha_2)^{pqr}_{~~~~l}F^{mnl}_{~~~~t}+(\beta_2)^{pqr}_{~~~~l}G^{mnl}_{~~~~t}
\nonumber \\
&&-F^{mnr}_{~~~~l}(\alpha_2)^{pql}_{~~~~t}+G^{mnr}_{~~~~l}(\beta_2)^{pql}_{~~~~t}=0~,
\label{YU2two}
\end{eqnarray}

\begin{eqnarray}
A^{pqr}_{~~~~s}B^{mns}_{~~~~t}-B^{pqr}_{~~~~~s}A^{mns}_{~~~~t}+A^{mnr}_{~~~~~s}B^{pqs}_{~~~~t}-B^{mnr}_{~~~~~s}A^{pqs}_{~~~~t}=0~,
\label{ZZone}
\end{eqnarray}
\begin{eqnarray}
&&A^{pqr}_{~~~~s}A^{mns}_{~~~~t}+B^{pqr}_{~~~~~s}B^{mns}_{~~~~t}+A^{mnr}_{~~~~~s}A^{pqs}_{~~~~t}+B^{mnr}_{~~~~~s}B^{pqs}_{~~~~t}
\nonumber \\
&&+4\sqrt{k}\delta^r_{~t}\bigg(-A^{mn(p}_{~~~~~~l}d^{q)l}+B^{mn(p}_{~~~~~~~l}c^{q)l}\bigg)=0~,
\label{ZZtwo}
\end{eqnarray}

\begin{eqnarray}
F^{pqr}_{~~~~s}B^{mns}_{~~~~t}-G^{pqr}_{~~~~~s}A^{mns}_{~~~~t}+A^{mnr}_{~~~~~s}G^{pqs}_{~~~~t}-B^{mnr}_{~~~~~s}F^{pqs}_{~~~~t}=0~,
\label{ZYone}
\end{eqnarray}
\begin{eqnarray}
&&F^{pqr}_{~~~~s}A^{mns}_{~~~~t}+G^{pqr}_{~~~~~s}B^{mns}_{~~~~t}+A^{mnr}_{~~~~~s}F^{pqs}_{~~~~t}+B^{mnr}_{~~~~~s}G^{pqs}_{~~~~t}
\nonumber \\
&&-4\sqrt{k}\delta^r_{~t}\bigg(A^{mn[p}_{~~~~~~l}c^{q]l}+B^{mn[p}_{~~~~~~~l}d^{q]l}\bigg)=0~,
\label{ZYtwo}
\end{eqnarray}

\begin{eqnarray}
F^{pqr}_{~~~~s}G^{mns}_{~~~~t}-G^{pqr}_{~~~~~s}F^{mns}_{~~~~t}+F^{mnr}_{~~~~~s}G^{pqs}_{~~~~t}-G^{mnr}_{~~~~~s}F^{pqs}_{~~~~t}=0~,
\label{YYone}
\end{eqnarray}
\begin{eqnarray}
&&F^{pqr}_{~~~~s}F^{mns}_{~~~~t}+G^{pqr}_{~~~~~s}G^{mns}_{~~~~t}+F^{mnr}_{~~~~~s}F^{pqs}_{~~~~t}+G^{mnr}_{~~~~~s}G^{pqs}_{~~~~t}
\nonumber \\
&&-4\sqrt{k}\delta^r_{~t}\bigg(F^{mn[p}_{~~~~~~l}c^{q]l}+G^{mn[p}_{~~~~~~~l}d^{q]l}\bigg)=0~,
\label{YYtwo}
\end{eqnarray}

\begin{eqnarray}
&&(\alpha_1)^{pqr}_{~~~~l}(\alpha_1)^{mnl}_{~~~~t}-(\beta_1)^{pqr}_{~~~~l}(\beta_1)^{mnl}_{~~~~t}-(\alpha_1)^{mnr}_{~~~~l}(\alpha_1)^{pql}_{~~~~t}+(\beta_1)^{mnr}_{~~~~l}(\beta_1)^{pql}_{~~~~t}
\nonumber \\
&&+2(\alpha_1)^{mn[p}_{~~~~~~l}(\alpha_1)^{q]lr}_{~~~~t}+2(\beta_1)^{mn[p}_{~~~~~~l}(\alpha_2)^{q]lr}_{~~~~t}=0~,
\label{u1u1uno}
\end{eqnarray}

\begin{eqnarray}
&&(\alpha_1)^{pqr}_{~~~~l}(\beta_1)^{mnl}_{~~~~t}+(\beta_1)^{pqr}_{~~~~l}(\alpha_1)^{mnl}_{~~~~t}-(\alpha_1)^{mnr}_{~~~~l}(\beta_1)^{pql}_{~~~~t}-(\beta_1)^{mnr}_{~~~~l}(\alpha_1)^{pql}_{~~~~t}
\nonumber \\
&&+2(\alpha_1)^{mn[p}_{~~~~~~l}(\beta_1)^{q]lr}_{~~~~t}+2(\beta_1)^{mn[p}_{~~~~~~l}(\beta_2)^{q]lr}_{~~~~t}=0~,
\label{u1u1due}
\end{eqnarray}

\begin{eqnarray}
&&(\alpha_2)^{pqr}_{~~~~l}(\alpha_1)^{mnl}_{~~~~t}-(\beta_2)^{pqr}_{~~~~l}(\beta_1)^{mnl}_{~~~~t}-(\alpha_1)^{mnr}_{~~~~l}(\alpha_2)^{pql}_{~~~~t}+(\beta_1)^{mnr}_{~~~~l}(\beta_2)^{pql}_{~~~~t}
\nonumber \\
&&+2(\alpha_1)^{mn[p}_{~~~~~~l}(\alpha_2)^{q]lr}_{~~~~t}-2(\beta_1)^{mn[p}_{~~~~~l}(\alpha_1)^{q]lr}_{~~~~t}=0~,
\label{u1u2uno}
\end{eqnarray}

\begin{eqnarray}
&&(\alpha_2)^{pqr}_{~~~~l}(\beta_1)^{mnl}_{~~~~t}+(\beta_2)^{pqr}_{~~~~l}(\alpha_1)^{mnl}_{~~~~t}-(\alpha_1)^{mnr}_{~~~~l}(\beta_2)^{pql}_{~~~~t}-(\beta_1)^{mnr}_{~~~~l}(\alpha_2)^{pql}_{~~~~t}
\nonumber \\
&&+2(\alpha_1)^{mn[p}_{~~~~~~l}(\beta_2)^{q]lr}_{~~~~t}-2(\beta_1)^{mn[p}_{~~~~~~l}(\beta_1)^{q]lr}_{~~~~t}=0~,
\label{u1u2due}
\end{eqnarray}

\begin{eqnarray}
&&(\alpha_2)^{pqr}_{~~~~l}(\alpha_2)^{mnl}_{~~~~t}-(\beta_2)^{pqr}_{~~~~l}(\beta_2)^{mnl}_{~~~~t}-(\alpha_2)^{mnr}_{~~~~l}(\alpha_2)^{pql}_{~~~~t}+(\beta_2)^{mnr}_{~~~~l}(\beta_2)^{pql}_{~~~~t}
\nonumber \\
&&+2(\alpha_2)^{mn[p}_{~~~~~~l}(\alpha_2)^{q]lr}_{~~~~t}-2(\beta_2)^{mn[p}_{~~~~~~l}(\alpha_1)^{q]lr}_{~~~~t}=0~,
\label{u2u2uno}
\end{eqnarray}

\begin{eqnarray}
&&(\alpha_2)^{pqr}_{~~~~l}(\beta_2)^{mnl}_{~~~~t}+(\beta_2)^{pqr}_{~~~~l}(\alpha_2)^{mnl}_{~~~~t}-(\alpha_2)^{mnr}_{~~~~l}(\beta_2)^{pql}_{~~~~t}-(\beta_2)^{mnr}_{~~~~l}(\alpha_2)^{pql}_{~~~~t}
\nonumber \\
&&+2(\alpha_2)^{mn[p}_{~~~~~~l}(\beta_2)^{q]lr}_{~~~~t}-2(\beta_2)^{mn[p}_{~~~~~~l}(\beta_1)^{q]lr}_{~~~~t}=0~,
\label{u2u2due}
\end{eqnarray}

\item{$(\mathfrak{g}_0,\mathfrak{g}_1,\mathfrak{g}_1)$:} the non-trivial conditions are\footnote{Note that the commutators of the internal Killing vectors \eqref{U2} are fixed by the super-Jacobi identities via \eqref{commUS1}.}

\begin{eqnarray}
&&[U_1^{mn},U_1^{rs}]=-2(\alpha_1)^{mn[r}_{~~~~~l}(U_1)^{s]l}-2(\beta_1)^{mn[r}_{~~~~~l}(U_2)^{s]l}
\nonumber \\
&&[U_2^{mn},U_2^{rs}]=-2(\alpha_2)^{mn[r}_{~~~~~l}(U_2)^{s]l}+2(\beta_2)^{mn[r}_{~~~~~l}(U_1)^{s]l}
\nonumber \\
&&[U_1^{mn},U_2^{rs}]=-2(\alpha_1)^{mn[r}_{~~~~~l}(U_2)^{s]l}+2(\beta_1)^{mn[r}_{~~~~~l}(U_1)^{s]l} 
\nonumber \\
&&(\alpha_2)^{rs[m}_{~~~~~l}(U_1)^{n]l}+(\beta_2)^{rs[m}_{~~~~~l}(U_2)^{n]l}=-(\alpha_1)^{mn[r}_{~~~~~l}(U_2)^{s]l}+(\beta_1)^{mn[r}_{~~~~~l}(U_1)^{s]l}
\nonumber \\
\label{commUS1}
\end{eqnarray}

\begin{eqnarray}
&&2(\alpha_1)^{rs(m}_{~~~~~l}f^{n)l}-2(\beta_1)^{rs(m}_{~~~~~l}g^{n)l}=A^{mn[r}_{~~~~~l}f^{s]l}-B^{mn[r}_{~~~~~l}g^{s]l}
\nonumber \\
&&2(\alpha_2)^{rs(m}_{~~~~~l}f^{n)l}-2(\beta_2)^{rs(m}_{~~~~~l}g^{n)l}=A^{mn[r}_{~~~~~l}g^{l]s}+B^{mn[r}_{~~~~~l}f^{s]l}
\nonumber \\
&&-2(\alpha_1)^{rs[m}_{~~~~~l}g^{n]l}-2(\beta_1)^{rs[m}_{~~~~~l}f^{n]l}=F^{mn[r}_{~~~~~l}f^{s]l}-G^{mn[r}_{~~~~~l}g^{s]l}
\nonumber \\
&&-2(\alpha_2)^{rs[m}_{~~~~~l}g^{n]l}-2(\beta_2)^{rs[m}_{~~~~~l}f^{n]l}=F^{mn[r}_{~~~~~l}g^{s]l}+G^{mn[r}_{~~~~~l}f^{s]l}
\nonumber \\
\label{fing0g0g1tre}
\end{eqnarray}

\begin{eqnarray}
&&[Z^{rs},Z^{pq}]=\frac{1}{4}B^{rs(p}_{~~~~l}(U_2)^{q)l}+\frac{1}{4}A^{rs(p}_{~~~~l}(U_1)^{q)l}
\nonumber \\
&&[Z^{rs},Y^{pq}]=\frac{1}{4}B^{rs[p}_{~~~~l}(U_1)^{q]l}-\frac{1}{4}A^{rs[p}_{~~~~l}(U_2)^{q]l}
\nonumber \\
&&[Y^{rs},Y^{pq}]=-\frac{1}{4}F^{rs[p}_{~~~~l}(U_2)^{q]l}+\frac{1}{4}G^{rs[p}_{~~~~l}(U_1)^{q]l}
\nonumber \\
&&G^{pq(r}_{~~~~~l}(U_2)^{s)l}+F^{pq(r}_{~~~~~l}(U_1)^{s)l}=-B^{rs[p}_{~~~~~l}(U_1)^{q]l}+A^{rs[p}_{~~~~~l}(U_2)^{q]l}
\nonumber \\
\label{fing0g0g1quattro}
\end{eqnarray}

\item{$(\mathfrak{g}_1,\mathfrak{g}_1,\mathfrak{g}_1)$}: the non-trivial conditions read

\begin{eqnarray}
&&2\sqrt{k}d^{rs}\delta^t_{q}-2\sqrt{k}d^{t(r}\delta^{s)}_{q}-(\alpha_1)^{t(rs)}_{~~~~~q}-(\beta_2)^{t(rs)}_{~~~~~q}=0
\nonumber \\
&&A^{rst}_{~~~q}-A^{t(rs)}_{~~~~q}+G^{t(rs)}_{~~~~q}-(\alpha_1)^{t(rs)}_{~~~~~q}+(\beta_2)^{t(rs)}_{~~~~~q}=0
\nonumber \\
&&B^{rst}_{~~~q}-B^{t(rs)}_{~~~~q}-F^{t(rs)}_{~~~~q}-(\alpha_2)^{t(rs)}_{~~~~~q}-(\beta_1)^{t(rs)}_{~~~~~q}=0
\nonumber \\
&&-2\sqrt{k}c^{rs}\delta^t_{q}+2\sqrt{k}c^{t(r}\delta^{s)}_{q}-(\alpha_2)^{t(rs)}_{~~~~~q}+(\beta_1)^{t(rs)}_{~~~~~q}=0
\nonumber \\
&&-G^{rst}_{~~~~q}-G^{t[rs]}_{~~~~q}+A^{t[rs]}_{~~~~q}-(\alpha_1)^{t[rs]}_{~~~~~q}+(\beta_2)^{t[rs]}_{~~~~~q}=0
\nonumber \\
&&-F^{rst}_{~~~~q}-F^{t[rs]}_{~~~~q}-B^{t[rs]}_{~~~~q}+(\beta_1)^{t[rs]}_{~~~~~q}+(\alpha_2)^{t[rs]}_{~~~~~q}=0~.
\label{g1g1g1sei}
\end{eqnarray}
\end{itemize}

\section*{Acknowledgments}

DF is partially supported by the STFC DTP Grant ST/S505742. DF would like to thank Jan Gutowski for useful discussions.

\section*{Data Management}

No additional research data beyond the data presented and cited in this work are needed to validate the research findings in this work.

\end{document}